\documentclass[useAMS,usenatbib]{mn2e}
\usepackage{amsmath}
\usepackage{graphicx}
\usepackage{epstopdf}
\usepackage{amssymb,latexsym}
\usepackage{multicol}
\usepackage{color}

\begin{document}

\title[Young star clusters in starburst rings]{Young star clusters in circumnuclear starburst rings}

\author[Richard de Grijs et al.]{Richard de Grijs$^{1,2,3}$, Chao
  Ma$^{1,2}$, Siyao Jia$^{2,4}$, Luis C. Ho$^{1,2}$ and Peter
  Anders$^5$\\
$^1$Kavli Institute for Astronomy \& Astrophysics, Peking University,
  Yi He Yuan Lu 5, Hai Dian District, Beijing 100871, China;\\
  grijs@pku.edu.cn\\
$^2$Department of Astronomy, Peking University, Yi He Yuan Lu 5, Hai
  Dian District, Beijing 100871, China\\
$^3$International Space Science Institute--Beijing, 1 Nanertiao,
  Zhongguancun, Hai Dian District, Beijing 100190, China\\
$^4$Department of Astronomy, University of California, Berkeley, CA
  94720, USA\\
$^5$Key Laboratory for Optical Astronomy, National Astronomical
  Observatories, Chinese Academy of Sciences, 20A Datun Road,\\ Chaoyang
  District, Beijing 100012, China\\ 
}

\date{xxx}

\pagerange{\pageref{firstpage}--\pageref{lastpage}} \pubyear{2016}
\label{firstpage}

\maketitle

\begin{abstract} 
We analyse the cluster luminosity functions (CLFs) of the youngest
star clusters in three galaxies exhibiting prominent circumnuclear
starburst rings. We focus specifically on NGC 1512 and NGC 6951, for
which we have access to H$\alpha$ data that allow us to unambiguously
identify the youngest sample clusters. To place our results on a firm
statistical footing, we first explore in detail a number of important
technical issues affecting the process from converting the
observational data into the spectral-energy distributions of the
objects in our final catalogues. The CLFs of the young clusters in
both galaxies exhibit approximate power-law behaviour down to the 90
per cent observational completeness limits, thus showing that star
cluster formation in the violent environments of starburst rings
appears to proceed similarly as that elsewhere in the local
Universe. We discuss this result in the context of the density of the
interstellar medium in our starburst-ring galaxies.
\end{abstract}

\section{Introduction}

Over the past decades, the prevalence in starburst environments of
luminous, young massive star clusters (YMCs) has become well
established, both based on ground-based data (e.g., van den Bergh
1971; Schweizer 1982; Arp \& Sandage 1985; Lutz 1991) and by
exploiting high-resolution {\sl Hubble Space Telescope} ({\sl HST})
observations (e.g., Holtzman et al. 1992; de Grijs et al. 2003a,b,
2013a; Whitmore 2003; Larsen et al. 2006). The majority of
extragalactic starburst environments host unresolved clusters, access
to whose individual member stars is, however, impossible with current
instrumentation. Consequently, the star cluster luminosity (mass)
function (CLF, CMF), i.e., the number of clusters per unit cluster
luminosity (mass), is among the most important diagnostics for our
understanding of the evolution of star cluster systems as a whole. The
CLFs of young star cluster systems have been studied extensively in
interacting and starburst galaxies, and they are often compared with
their counterparts in the discs of `normal', quiescent galaxies.

Until approximately a decade ago, a fierce debate about the evolution
of CLFs and CMFs raged in the astrophysics community (for a review,
see de Grijs \& Parmentier 2007). One school of thought (e.g.,
Whitmore \& Schweizer 1995; Whitmore et al. 1999; Zhang \& Fall 1999;
Mengel et al. 2005) advocated that the CLF (CMF) at the time of
cluster formation was a power-law function of the form $N(L_V) \propto
L_V^{-\alpha}$ or, equivalently, $N(M_{\rm cl}) \propto M_{\rm
  cl}^{-\alpha}$, where $L_V$ and $M_{\rm cl}$ are the clusters'
$V$-band luminosities and masses, respectively, and $\alpha \approx 2$
(e.g., de Grijs et al. 2003b; Portegies Zwart et al. 2010; Fall \&
Chandar 2012). This initial power-law CLF (CMF) would gradually
transform into a Gaussian or `lognormal' distribution, to eventually
resemble the near-universal distributions of the old globular cluster
systems found throughout the local Universe (e.g., Fall \& Zhang 2001;
Fall 2006). The competing scenario started from an initially lognormal
CLF (CMF), which would retain its lognormal shape over the entire
evolutionary time-scale of a given star cluster system, although the
peak luminosity (mass) would gradually be reduced (e.g., Vesperini
1998; Anders et al. 2007).

This debate was predominantly fuelled by a number of potentially
unphysical assumptions underlying the main study supporting the
initial power-law CMF arguments at that time (Fall \& Zhang 2001). In
addition, the observational data available around the turn of the
century were insufficiently deep to allow one to reach unequivocally
below the postulated peak of the CLF in what became the poster-child
star cluster system in this field, the Antennae system (NGC 4038/9;
e.g., Anders et al. 2007). In the mean time, deeper observations of
both the Antennae system (e.g., Whitmore et al. 2010) as well as of
the young cluster systems in both the Small and Large Magellanic
Clouds (e.g., de Grijs \& Goodwin 2008; de Grijs et al. 2013b) have
clearly shown that the CLFs of the youngest (disc) star cluster
systems are well-described by power-law shapes down to the lowest
luminosities attainable, corresponding to masses well below $M_{\rm
  cl} = 10^4$ M$_\odot$. Theoretical efforts (e.g., Fall 2006) have
also adequately and convincingly addressed the earlier criticism
levelled at Fall \& Zhang's (2001) seminal study. The community's
commonly accepted point of view has thus become that star clusters, in
both galactic discs and starburst regions, usually form following a
power-law CMF (CLF), which is gradually transformed into a
lognormal-like distribution owing to both internal and external
perturbations and two-body relaxation processes, which operate on
billion-year time-scales.

In this context, our recent discovery (de Grijs \& Anders 2012) of a
lognormal-like CMF for star clusters as young as $\sim 10$ Myr in the
circumnuclear starburst ring of the spiral galaxy NGC 7742 came as a
surprise. Neither theoretical predictions nor prior observational
evidence indicated that initial power-law CLFs/CMFs could transform
into lognormal distributions on time-scales as short as a few $\times
10^7$ yr, yet that was what we inferred to have occurred. This led us
to propose that the physical conditions governing star cluster
evolution and possibly also their formation may be different in the
dense, highly complex starburst-ring environments from those both in
quiescent galaxy discs and in interacting and starburst galaxies.

However, few studies to date have specifically addressed the CLFs
pertaining to circumnuclear starburst rings or their evolutionary
scenarios. Yet, many galaxies feature remarkable circumnuclear
starburst rings, which are associated with high star-formation
rates. Two scenarios have been proposed to explain their formation:
(i) as a consequence of bar-driven gas inflow and dynamical resonances
in the bar, inflowing gas as well as stars accumulate in a ring which
connects the two inner Lindblad resonances at the ends of the bars
(e.g., Romero-G\'{o}mez et al. 2006; Athanassoula et al. 2010;
Athanassoula 2012); or (ii) they may have formed owing to the
centrifugal barrier encountered by gas migrating to the inner regions
of the galaxy (Kim et al. 2012). The high-density environments in
these rings make them ideal locations to harbour large numbers of
YMCs, much more so than in galaxy centres (e.g., Miocchi et
al. 2006). Indeed, based on high-resolution {\sl HST} images, numerous
young (a few $\times 10^7$ yr-old) and intermediate-age (a few
Gyr-old) star clusters have been uncovered in these structures (e.g.,
Barth et al. 1995; Maoz et al. 1996, 2001; Ho 1997; Buta et al. 2000;
de Grijs et al. 2003a,b; Mazzuca et al. 2008; de Grijs \& Anders 2012;
Hsieh et al. 2012; van der Laan et al. 2013, henceforth vdL13).

In this paper, we explore the CLFs of the youngest star clusters in
three well-observed galaxies with large amounts of ancillary data
which exhibit prominent circumnuclear starburst rings. We start by
focussing on the starburst-ring cluster population in the spiral
galaxy NGC 6951, an SAB(rs)bc galaxy with a well-known and
well-studied ring: see Barth et al. (1995) for an early {\sl
  HST}-based cluster study, as well as the review by vdL13 (and
references therein). We will also examine in detail the CLF of the
clusters in the remarkably luminous circumnuclear starburst ring in
NGC 1512. NGC 1512 is an SB(r)a galaxy at a distance of $\sim$10 Mpc
(e.g., Maoz et al. 2001). We first explore in detail a number of
technical issues: see Section 2. In Section 3, we address the shapes
of the CLFs in our two main sample galaxies. Finally, we place the
results obtained for the individual ring populations in the broader
context of the field of star cluster evolution in Section 5.

\section{From observational data to star cluster photometry}

\subsection{Data sets}

Our perusal of the {\sl HST} Data Archive showed that NGC 6951 has
been observed by a number of investigators using a range of
cameras. Table \ref{data.tab} (top) includes details of the full data
set acquired from the Hubble Legacy Archive
(HLA).\footnote{http://hla.stsci.edu/hlaview.html} The imaging
observations, already processed and calibrated by the HLA pipeline
reduction software, span the entire optical/near-infrared wavelength
range, from the bluest F330W filter observed with the Advanced Camera
for Surveys (ACS)--High-Resolution Camera (HRC) to the near-infrared
F160W filter obtained with the Near-Infrared Camera and Multi-Object
Spectrometer's (NICMOS) Camera 2 (NIC2).

\begin{table*}
\begin{center}
  \caption{Detailed observational characteristics of the adopted {\sl
      HST} data sets.}
  \label{data.tab}
  \begin{tabular}{lllcc}
  \hline
   Filter   & Proposal ID/PI & Camera & Exposure time (s) & PHOTFLAM \\
 \hline
\multicolumn{5}{c}{NGC 6951}\\
\cline{1-5}
F330W ($U$) & GO-9379/Schmitt   & ACS/HRC     & 1200   & $2.2367671\times10^{-18}$\\
F547M ($V$) & GO-5419/Sargent   & WFPC2/PC    & 300    & $1.0802045\times10^{-18}$\\
F606W ($R$) & GO-8597/Regan     & WFPC2/PC    & 560    & $2.6681081\times10^{-19}$\\
F658N (H$\alpha$) & GO-9788/Ho  & ACS/WFC     & 700    & $1.9614893\times10^{-18}$\\
F814W ($I$) & GO-9788/Ho        & ACS/WFC     & 120    & $6.9504181\times10^{-20}$\\
F110W ($YJ$)& GO-7331/Stiavelli & NICMOS/NIC2 & 256    & $4.3320170\times10^{-19}$\\
F160W ($H$) & GO-7330/Mulchaey  & NICMOS/NIC2 & 320    & $2.3600094\times10^{-19}$\\
\hline
\multicolumn{5}{c}{NGC 1512}\\
\cline{1-5}
F336W ($U$) & GO-13364/Calzetti & WFC3        & 1107   & $1.31683050\times10^{-18}$\\
F438W ($B$) & GO-13364/Calzetti & WFC3        &  953   & $6.91138715\times10^{-19}$\\
F555W ($V$) & GO-13364/Calzetti & WFC3        & 1131   & $1.87652800\times10^{-19}$\\
F658N (H$\alpha$) & GO-6738/Filippenko & WFPC2/PC& 5200& $1.45442135\times10^{-17}$\\
F814W ($I$) & GO-13364/Calzetti & WFC3        &  977   & $1.53047990\times10^{-19}$\\
\hline
\end{tabular}
\end{center}
\end{table*}

We rotated, aligned and cropped the combined images in all passbands
to a common orientation, pixel scale and field of view, respectively,
using standard {\sc iraf/stsdas} routines\footnote{The Image Reduction
  and Analysis Facility ({\sc iraf}) is distributed by the National
  Optical Astronomy Observatories, which is operated by the
  Association of Universities for Research in Astronomy, Inc., under
  cooperative agreement with the U.S. National Science
  Foundation. {\sc stsdas}, the Space Telescope Science Data Analysis
  System, contains tasks complementary to the existing {\sc iraf}
  tasks. We used Version 3.6 (November 2006) for the data reduction
  performed in this paper.} and adopting the ACS/WFC footprint as our
basis. The final image set consists of images in seven filters with a
pixel size of 0.05 arcsec, covering 999$\times$976 pixels in right
ascencion (R.A.) and declination (Dec), and centred on the galaxy
centre (assumed to coincide with the nuclear star cluster),
R.A. (J2000) = 20$^{\rm h}$ 37$^{\rm m}$ 14.13$^{\rm s}$, Dec (J2000)
= 66$^{\circ}$ 06$'$ 20.2$''$. This centre position is, to within a
fraction of an arcsecond, identical to that listed in the Two Micron
All-Sky Survey (2MASS; Skrutskie et al. 2006).

We also selected pre-processed, high-resolution images of NGC 1512
from the HLA, ensuring that they covered the longest possible
wavelength range. The most suitable data-set combination was observed
with the Wide Field Camera 3 (WFC3; pixel size $\sim 0.04$ arcsec) in
the F336W, F438W, F555W and F814W filters: see Table \ref{data.tab}
(bottom). All four exposures shared the same orientation, pixel scale
and field of view, covering the entire galaxy. Since the circumnuclear
starburst ring represents a small fraction of the full image, for
convenience we trimmed the ring portion in the original exposures to
yield a final standard science image of $585 \times 585$
pixels$^2$. In addition, we also explored the usefulness of the WFC3
ultraviolet F275W image, but this exposure was discarded because of
large photometric uncertainties.

\subsection{Object selection}

\begin{figure}
\begin{center}
\includegraphics[width=\columnwidth]{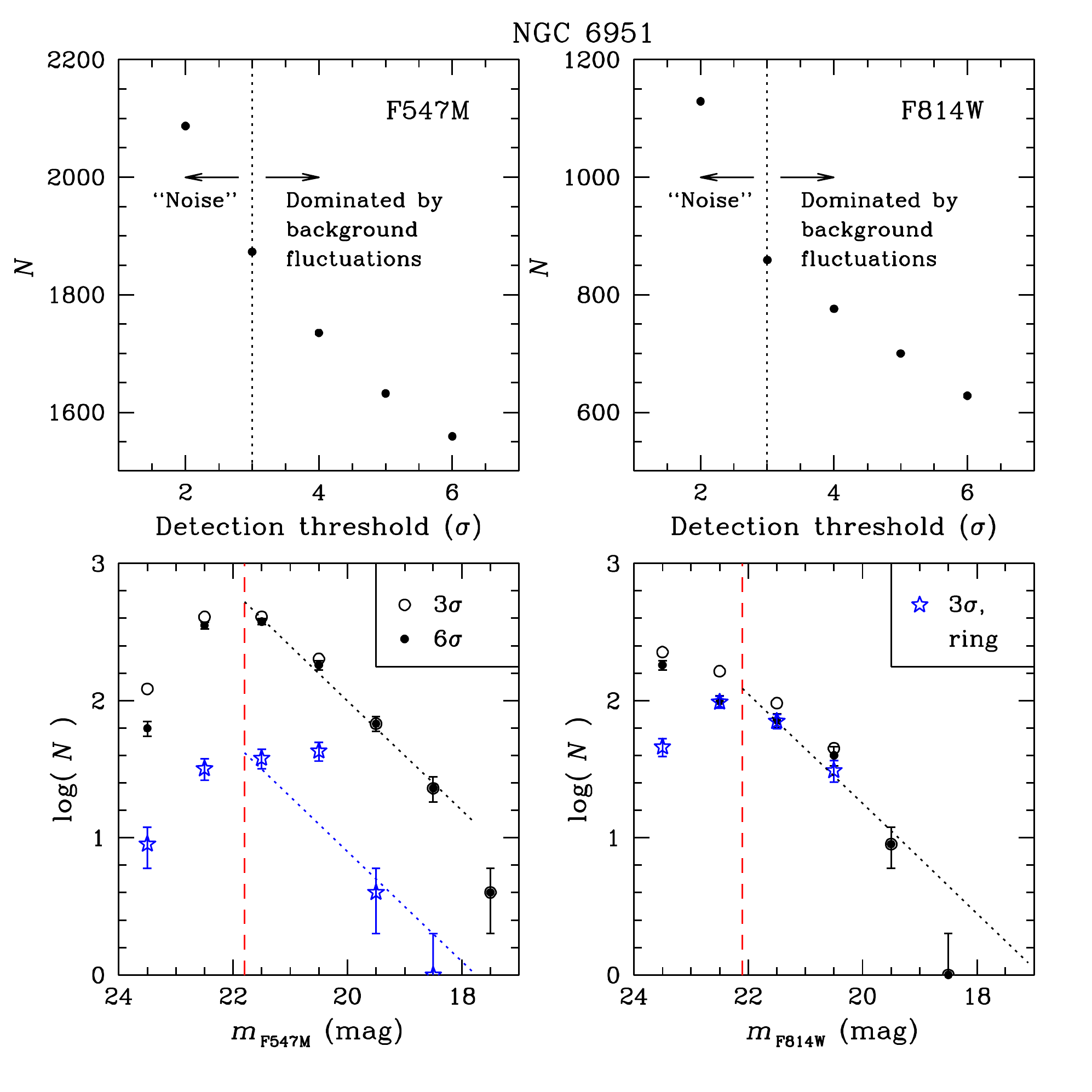}
\caption{Top: Numbers of detections as a function of detection
  threshold, in units of the standard deviation of the background
  field, $\sigma$, for the F547M and F814W images centred on NGC
  6951. The noise- and background fluctuation-dominated ranges are
  indicated; `background fluctuations' also include a minority of real
  sources. Bottom: Luminosity functions of the objects detected in the
  F547M and F814W images (for detections with maximum photometric
  uncertainties of 0.5 mag in the relevant filter) prior to
  application of any additional selection criteria. Shown are the
  results for the full field of view (open circles and solid bullets)
  as well as for the ring area only (blue asterisks), for different
  detection thresholds (see figure legend). The canonical $\alpha = 2$
  power-law slopes are indicated by dotted lines (which are not fits
  to the data); the relevant 90 per cent completeness limits
  (applicable to the full fields of view of our science images) are
  shown as the vertical red dashed lines. The numbers of detections
  indicated along the vertical axes of the bottom panels represent the
  numbers of detections per magnitude bin. We adopt this convention
  throughout this paper.}
\label{fig0}
\end{center}
\end{figure}

Following the same steps as in our previously established image
reduction and analysis protocol (e.g., de Grijs et al. 2013a; Li et
al. 2015; and references therein), we ran our custom-written {\sc
  idl}\footnote{The Interactive Data Language ({\sc idl}) is licensed
  by Research Systems Inc., of Boulder, CO, USA.} source-finding
routines on two `middle-wavelength' images for each galaxy. For NGC
6951 we adopted the F547M and F814W observations, while for NGC 1512
we used the F438W and F555W images. We selected these
middle-wavelength images to ensure that their use would not directly
lead to rejection of extremely blue or extremely red objects from our
initial sample of cluster candidates. We based our assessment of the
number of detections as a function of detection threshold (for
details, see also Barker et al. 2008), in units of the
root-mean-square scatter in the background level, $\sigma_{\rm
  bg}$. For a given filter, the latter was determined based on the
standard deviation of the `counts' in an empty section of the science
frames, located as far away from the galaxy as possible. For both
galaxies we adopted minimum detection limits of 3$\sigma_{\rm bg}$ in
all four middle-wavelength images. This choice returned 1873 and 1129
detections for NGC 6951 in the F547M and F814W images,
respectively. For the NGC 1512 data, application of the 3$\sigma_{\rm
  bg}$ critical thresholds in the F438W and F555W bands resulted in
catalogues containing 2304 and 1834 detections, respectively.

This first selection step served, in essence, to remove false
detections owing to Poissonian shot noise from our initial catalogues;
the resulting catalogues contain both genuine objects and small-scale
background fluctuations, thus necessitating additional selection
steps. As an example, in Fig. \ref{fig0} (top panels) we show for the
NGC 6951 observations the number of detections as a function of
threshold in both the F547M and F814W images. Adopting a detection
threshold of 3$\sigma_{\rm bg}$, based on a change in slope of the
detection efficiency (i.e., the number of detections as a function of
minimum background level), ensures that we are conservative in
excluding real objects, so that this first selection step will not
unduly bias the final source counts.

The subsequent cross correlation of source positions, initially
allowing a positional mismatch of only 1 pixel in both spatial
dimensions, led to final source counts of 104 and 960 for NGC 6951 and
NGC 1512, respectively.\footnote{Two of us (R. d. G. and S. J. for NGC
  6951; C. M. and R. d. G. for NGC 1512) analysed the {\sl HST}
  archive data independently; we both found a very similar number of
  genuine cluster candidates.} As we will show in Section
\ref{biases.sec}, releasing this constraint, adjusted to 2- and
3-pixel positional mismatches in both directions, does not have a
significant effect on the number of objects retained for further
analysis. The reduction from the initial numbers of detections in the
individual passbands to our final source catalogues is
significant. However, as we will show below by using detection
threshold based on the intensity variations in the ring regions
instead of those in empty fields, the steps we have followed to obtain
our final samples are well-understood and justifiable: each step
approached the selection from a rather conservative angle so as not to
remove genuine objects inadvertently. This meant that our initial
source lists were dominated by small-scale background fluctuations
rather than real clusters. The final cross-correlation step (and, for
NGC 1512, the minimum size selection applied below) is the most
important procedure in whittling down the initial,
fluctuation-dominated catalogue of detections to genuine object lists.

\begin{figure}
\begin{center}
\includegraphics[width=0.72\columnwidth]{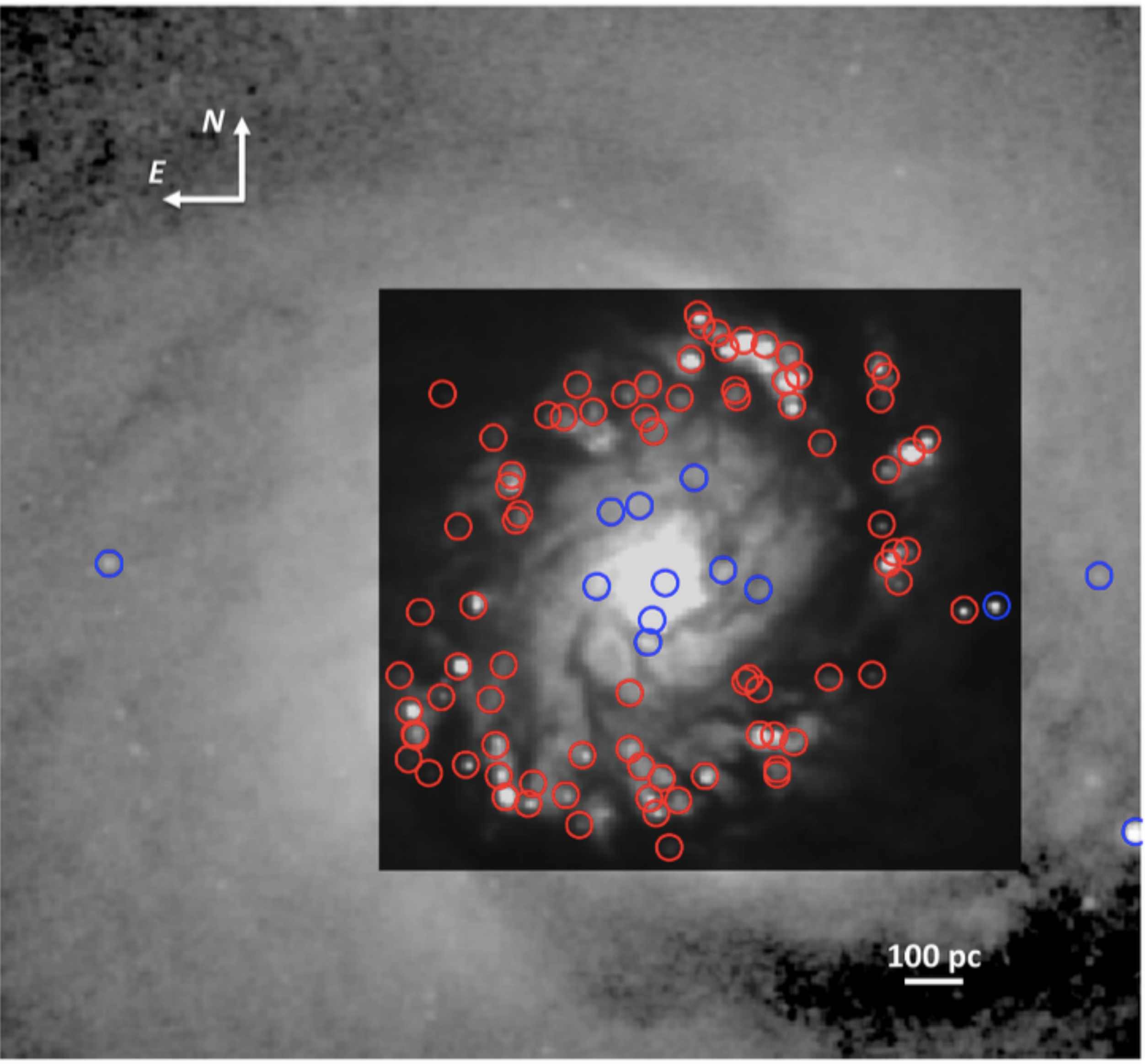}
\includegraphics[width=\columnwidth]{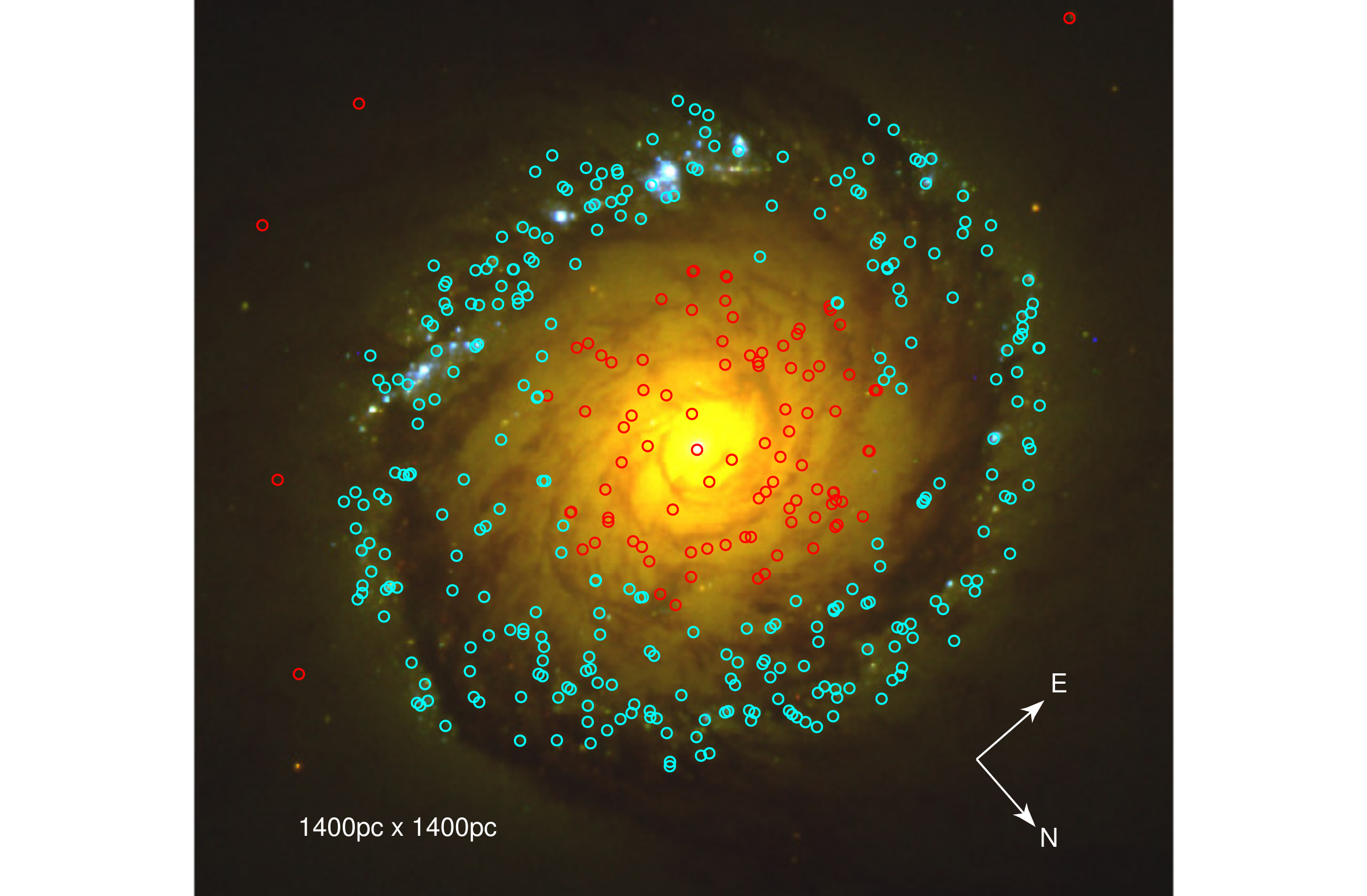}
\caption{Top: Inner $350 \times 350$ pixels$^2$ ($17.5 \times 17.5$
  arcsec$^2$) of our NGC 6951 observations as seen through the F606W
  filter. The star clusters that form part of the starburst ring and
  those not associated with the ring are shown with red and blue
  circles, respectively. For optimal contrast, the inner section of
  this panel has been displayed in linear intensity units, while the
  outer section is shown using a logarithmic scale. Bottom: Composite
  image of our target starburst ring area in the centre of NGC 1512
  ($1400 \times 1400$ pc$^2$), originally composed for a press release
  (http://spacetelescope.org/news/heic0106/). The cyan circles
  indicate the clusters associated with the ring, while their red
  counterparts are clusters in the galaxy's main disc.}
\label{newfig1}
\end{center}
\end{figure}

Of the 104 cluster candidates in NGC 6951, 82 are associated with the
circumnuclear starburst ring, which we broadly define as the dominant
feature at radii between 35 and 105 pixels, or $200 \le R \le 600$ pc
if we adopt an absolute distance modulus of $(m-M)_0 = 31.87 \pm 0.32$
mag (23.7 Mpc).\footnote{This value represents the geometric mean of
  11 individual distance measurements, 10 of which are based on
  distance estimates to supernova SN 2000E (NASA Extragalactic
  Database;
  http://ned.ipac.caltech.edu/cgi-bin/nDistance?name=NGC+6951). The
  median distance modulus to NGC 6951 based on these measurements is
  $(m-M)_0 = 31.84$ mag.} At this distance, 1 arcsec $\equiv 115$ pc,
while 1 ACS/WFC pixel $\equiv 6$ pc. Figure \ref{newfig1} (top)
displays the galaxy's inner $350 \times 350$ pixels$^2$ ($17.5 \times
17.5$ arcsec$^2$) as observed through the F606W filter, with the star
clusters that form part of the starburst ring and those not associated
with the ring indicated by red and blue circles, respectively. We
opted to show the F606W image instead of the F547M image on which our
cluster selection is partially based, because the former is much
deeper and shows more structure while covering a very similar
wavelength range; the latter image is additionally affected by
significant numbers of cosmic-ray hits (e.g., examine in detail fig. 4
of vdL13; residuals of their cosmic-ray removal appear as white areas
without background noise), none of which coincide with any of our
genuine star cluster candidates, however.

Because of the galaxy's closer proximity, for NGC 1512 we proceeded by
first applying a cluster size cut in order to remove stellar (point)
sources. To do so, we made use of a standard Gauss-fitting routine in
{\sc idl}, applied to all selected sources, to determine their sizes,
$\sigma_{\rm G}$. To define the minimum size for extended cluster
candidates, we generated artificial {\sl HST} point-spread functions
(PSFs) using the {\sc TinyTim} software package (Krist \& Hook
1997). The best {\it Gaussian} width resulting from this analysis was
$\sigma_{\rm G,PSF}=1.24$ pixels. Note that the {\sl HST} PSF is not a
two-dimensional (2D) Gaussian function. However, by consistently
applying computationally robust Gaussian fits to our 2D object
data,\footnote{Note that while realistic profiles can be fitted to the
  brighter objects, their complexity renders such fits to the fainter
  objects unstable but Gaussian fits turn out to be much more
  computationally robust and stable. Therefore, we opted to apply
  Gaussian fits to all candidate clusters so as to gain insights into
  their {\it relative} size distributions.} we can nevertheless
distinguish very well the more extended objects from their unresolved
counterparts. A side effect of this approach is that some real objects
will be returned with Gaussian sizes that are smaller than the {\sc
  TinyTim} $\sigma_{\rm G,PSF}$, simply because of the differences
between the actual {\sl HST} PSF and a smooth 2D Gaussian
distribution.

We hence adopted a conservative size-cut criterion so as not to reject
some marginally extended sources (i.e., $\sigma_{\rm G,min} \geq 1.14$
pixels = 2.74 pc). Note that 1 WFC3 pixel corresponds to roughly 2.4
pc in linear size, for an absolute distance modulus to NGC 1512 of
$(m-M)_0 = 30.48 \pm 0.25$ mag (13.0 Mpc).\footnote{This value
  represents the geometric mean of 10 individual distance measurements
  compiled in the NASA Extragalactic Database
  (http://ned.ipac.caltech.edu). The median distance modulus to NGC
  1512 based on these measurements is $(m-M)_0 = 30.57$ mag.} We
verified that our size cut, combined with our adopted minimum
brightness levels, is adequate to avoid individual stars, assuming a
typical stellar full width at half maximum of $\sim 2.2$ pixels
($\sigma \sim 0.93$ pixel) for WFC3/UVIS images (e.g., Calzetti et
al. 2015; based on the actual {\sl HST} profiles instead of Gaussian
fits). In fact, an increase of our adopted size cut will exclude most
low-mass clusters. This step left us with a sample of 469 cluster
candidates (as discussed for NGC 6951 in Section \ref{photom.sec}, we
also removed several sources with extremely large Gaussian widths,
$\sigma_{\rm G} \ge$ 43 pc). Figure \ref{newfig1} (bottom) shows the
spatial distribution of our resulting NGC 1512 cluster sample
overplotted on a colour composite image. The cluster candidates
identified by the cyan circles are located in the NGC 1512
circumnuclear starburst ring, which we delineated by an annular region
containing the main, conspicuous features at radii between 104 and 260
pixels ($288\leq R \leq 722$ pc), assuming an inclination angle of
30$\degr$ (Koribalski \& L\'opez-S\'anchez 2009).

Careful visual inspection of the original images and a large subset of
the initial detections thus obtained reveals that our initial
catalogue is dominated by background fluctuations rather than genuine
clusters. The cross-correlation step is highly efficient in removing
the majority of these background fluctuations from our final source
list; as such, we consider this a crucial selection step. To back up
this statement quantitatively, we also pursued an alternative
approach: instead of defining our detections in terms of their flux
above a threshold based on the Poissonian noise characteristics of
empty areas in our science frames, for both galaxies we determined the
standard deviations across the starburst ring in all intermediate-band
filters used for the source detections. These standard deviations
include the effects of shot noise and fluctuations in the background
flux. For NGC 6951, again applying a $3\sigma$ threshold, this
selection yields 1119 and 436 detections in the F547M and F814W
images, respectively. The corresponding numbers for NGC 1512, also
using a $3\sigma$ threshold, are 553 and 327 objects in the F438W and
F555W filters, respectively. Note that these detections still include
small-scale background fluctuations, so that application of a
cross-correlation step is just as crucial in this case. The latter
step yields final object samples containing 33 and 289 objects in NGC
6951 and NGC 1512, respectively; the minimum size criterion adopted
for NGC 1512 further reduces this latter number to 153 ring
clusters. These final numbers of clusters are indeed smaller than the
numbers in our object catalogues resulting from application of
detection thresholds based on the noise statistics in empty field
regions. Indeed, the objects included in those latter catalogues but
not in the catalogues resulting from the use of detection thresholds
based on the intensity fluctuations in the galaxies' ring regions
represent the fainter complement of the CLFs. Their numbers are
consistent with an approximately homologous cluster population
obeying power-law CLFs characterized by $\alpha \sim 2$.

\subsection{Our source selection criteria: effects of biases}
\label{biases.sec}

We will now address the effects of our source selection procedures to
provide sound evidence of the physical reality of the objects
considered here. In Fig. \ref{fig0} (bottom panels) we show the
luminosity functions\footnote{For the construction of the luminosity
  functions in Fig. \ref{fig0}, we adopted fixed source aperture radii
  of 3 pixels and sky annuli of 6--8 pixels, instead of the adaptive
  aperture sizes we advocate in Section \ref{photom.sec} for the more
  advanced analysis of our genuine cluster sample. In de Grijs et
  al. (2013a) we showed that this approach leads to statistically
  identical results for large numbers of sample objects, such as those
  considered here.} of the detections in NGC 6951 in both the F547M
and F814W images, prior to any positional cross matching. We show the
number of detections based on adoption of thresholds of 3$\sigma_{\rm
  bg}$ and 6$\sigma_{\rm bg}$ (open circles and solid bullets,
respectively) in our full fields of view -- and using detection
thresholds based on the noise statistics in `empty' regions on the
science frames -- as well as the detections in the galaxy's starburst
ring area only (blue asterisks), i.e., for $200 \le R \le 600$ pc or
$35 \le R \le 105$ pixels. The luminosity functions are composed of
detections with maximum photometric uncertainties of 0.5 mag in the
relevant filter. However, releasing this constraint does not cause any
appreciable differences, except in the lowest-luminosity bins in both
filters, i.e., well below our completeness limits.

\begin{figure}
\begin{center}
\includegraphics[width=\columnwidth]{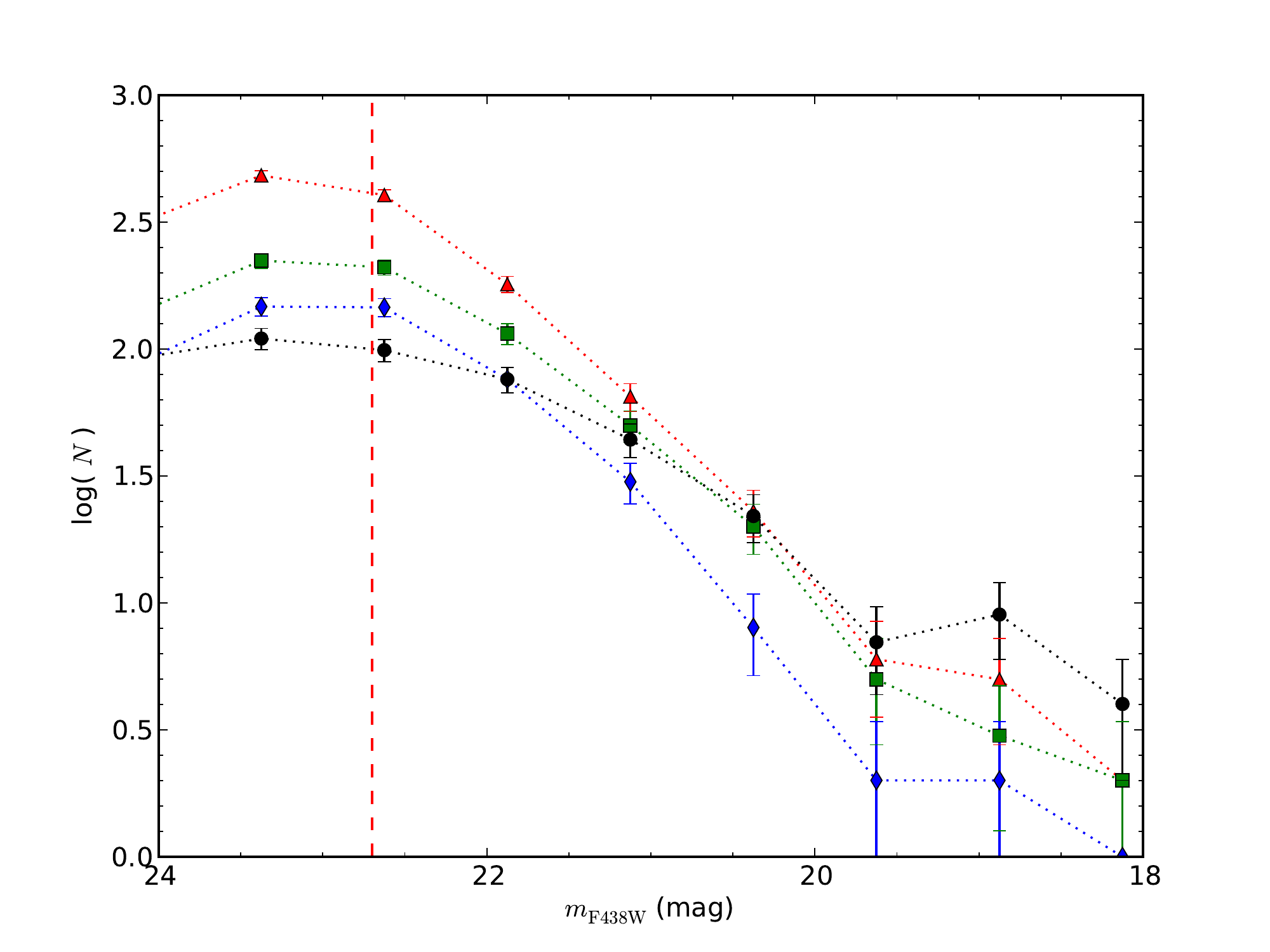}
\caption{Illustration of the effects of the different selection
  criteria applied to the NGC 1512 data to obtain our final sample of
  genuine cluster candidates. The error bars represent the Poissonian
  uncertainties. Red triangles: initial, single-passband luminosity
  function, using fixed apertures (source radii: 6 WFC3 pixels,
  background annuli: 7--10 pixels in radius). Green squares: CLF after
  F438W $\otimes$ F555W cross-correlation, using fixed apertures. Blue
  diamonds: CLF after application of a minimum $\sigma_{\rm G}$ cut,
  using fixed apertures. Black solid bullets: final CLF, using
  variable instead of fixed apertures. The vertical red dashed line
  indicates the 90 per cent completeness limit.}
\label{steps.fig}
\end{center}
\end{figure}

Next, we consider the effects of the different selection criteria used
to reach our final sample of genuine cluster candidates. We take the
NGC 1512 cluster population as an example and refer to
Fig. \ref{steps.fig} for illustrative purposes, where we show the NGC
1512 CLFs derived from our F438W photometry. The error bars represent
the Poissonian uncertainties. The initial, single-passband luminosity
function, using fixed source apertures of 6 WFC3 pixels and background
annuli spanning from 7 to 10 pixels in radii, is shown using red
triangles. In our next step, we cross-correlated the detections in the
F438W and F555W filters, yielding the CLF represented by green
squares; the apertures used for the corresponding photometry are
identical to those used for the initial luminosity function. The blue
diamonds trace the NGC 1512 CLF after our application of a minimum
size criterion in terms of the observational $\sigma_{\rm G}$, again
using the same photometric apertures. Finally, the black solid bullets
show the final CLF of genuine clusters, but using variable instead of
fixed apertures (for justification, see Section \ref{photom.sec}). It
is clear that the most significant change in the (C)LF shape occurs
when adjusting the aperture sizes used for the photometry. The blue
and black CLFs correspond to the same cluster sample, yet their shapes
are significantly different, in the sense that the number of bright
objects is much larger in the variable-aperture CLF than in the
fixed-aperture CLF. The reverse applies to the fainter objects.

\begin{figure}
\begin{center}
\includegraphics[width=\columnwidth]{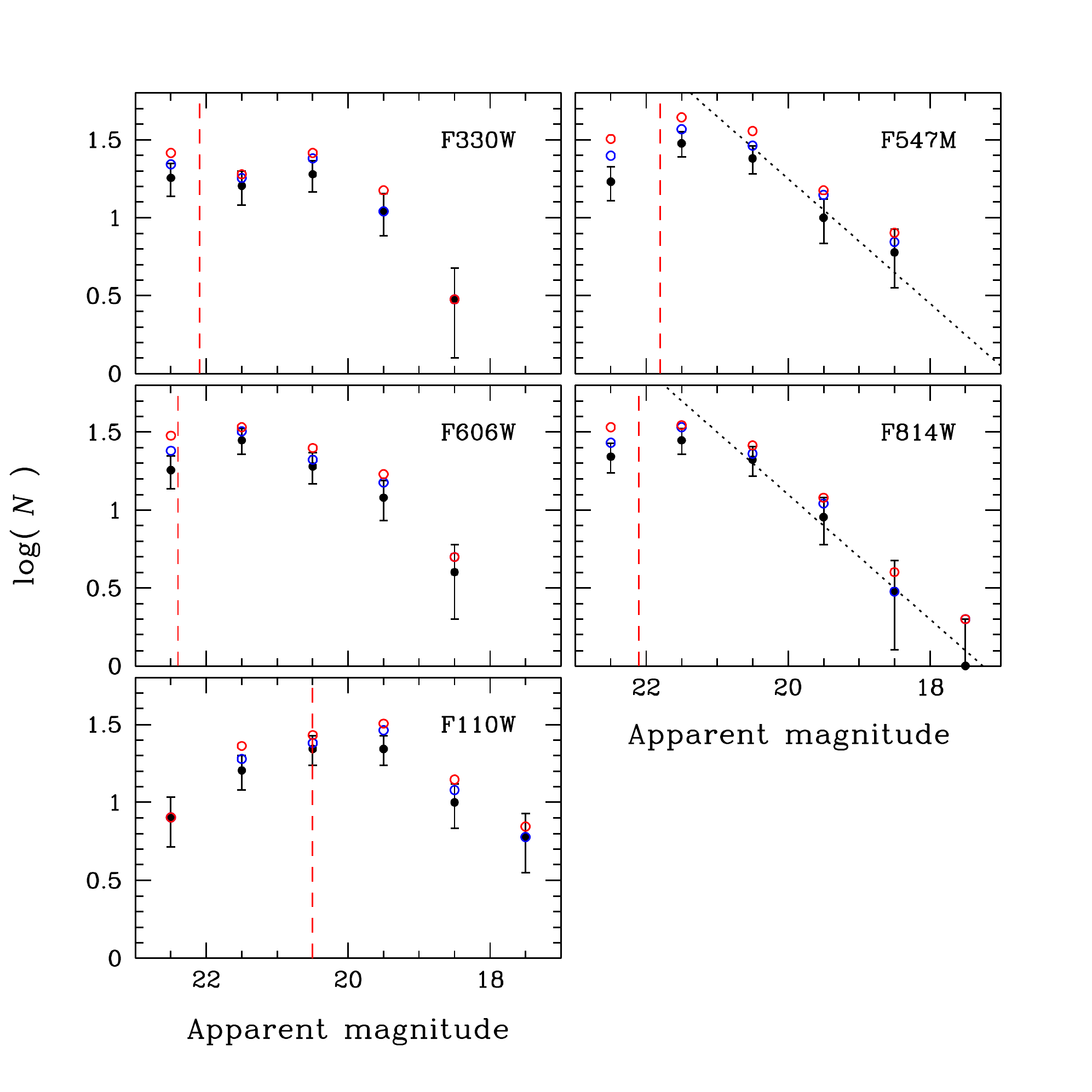}
\caption{Full NGC 6951 CLFs, based on application of size-dependent,
  variable apertures. The 90 per cent completeness levels are
  indicated by the vertical red dashed lines. Error bars denote
  Poissonian uncertainties. Black solid bullets, blue and red open
  cicles denote CLFs composed of clusters identified on the basis of
  maximum spatial differences of $1 \times 1$ pixel, $2 \times 2$
  pixels and $3 \times 3$ pixels in both spatial dimensions,
  respectively. For reasons of clarity, we have omitted the error bars
  pertaining to the CLFs resulting from application of our $2 \times
  2$- and $3 \times 3$-pixel selection criteria. The black dotted
  lines overplotted on the F547M and F814W CLFs represent canonical
  $\alpha = 2$ power laws.}
\label{mismatch.fig}
\end{center}
\end{figure}

We suggest that this effect is caused by the extended nature of our
sample objects (see Section \ref{photom.sec}). This is supported by a
comparison of the initial F547M and F814W luminosity functions of all
detections shown in the bottom panels of Fig. \ref{fig0} with the
right-hand panels of Fig. \ref{mismatch.fig}. We will discuss
Fig. \ref{mismatch.fig} in more detail below, but for the present
purpose it is sufficient to realize that the CLF slopes between
Fig. \ref{fig0} and Fig. \ref{mismatch.fig} are, in essence, the same
for luminosities well away from the 90 per cent completeness limits in
both filters (see Section \ref{compl.sec} below). We remind the reader
that we did not apply a minimum size criterion to the NGC 6951 cluster
sample because of the galaxy's location at a much greater distance
than that of NGC 1512. This fortuitously renders the majority of our
NGC 6951 sample clusters point sources for all practical purposes.

Finally, we explore the impact, if any, of having selected objects
found within 1 pixel in both $X$ and $Y$ between our two
middle-wavelength images. We consider the cross correlation of the
detections in two filters to generate our initial master list an
essential step to ensure that we are eventually dealing with genuine
clusters rather than noise peaks or other artefacts. However, although
the completeness levels in both filters are reasonably close, one
could envision a situation in which a cluster's SED is sufficiently
steep over this narrow wavelength range that it may be detectable in
one filter but fall below the detection threshold in the other. Since
such conditions would apply to luminosities close to the detection
threshold in either filter, this might have a measurable effect on the
resulting CLFs close to the canonical completeness limits.

Instead of allowing only a 1-pixel mismatch in both spatial
dimensions, we repeated our sample selection for both NGC 6951 and NGC
1512 adopting both 2- and 3-pixel mismatches (i.e., maximum spatial
differences of 2.8 pixels and 4.2 pixels, respectively), given that at
the faintest levels close to the detection limit stochastic dither
might affect an object's centre position. Compared with our 104 NGC
6951 sample objects based on the 1-pixel criterion, the 2- and 3-pixel
selection criteria resulted in samples of 131 and 158 candidate
clusters. The resulting CLFs are shown in
Fig. \ref{mismatch.fig}. Clearly, the more relaxed
cross-identification criteria did not lead to significantly different
CLFs.

Similarly, we reran our source selection routines on the NGC 1512 data
using the more relaxed 2- and 3-pixel selection restrictions to see
how the resulting shapes of the CLFs depend on the selection
criteria. Our $2 \times 2$- and $3 \times 3$-pixel selection criteria
resulted in samples of 1228 and 1440 cluster candidates, compared with
960 real objects if we adopt the original, $1 \times 1$-pixel
selection criterion. Again, the more relaxed cross-identication
criteria did not cause signicant deviations from the originally
derived CLFs.

\subsection{Cluster photometry}
\label{photom.sec}

We next obtained the broad-band SEDs for all 104 clusters identified
in the NGC 6951 central galaxy field using two different
approaches. For comparison with previous results, we first determined
the cluster photometry based on 3-pixel source apertures (based on
inspection of the radial profiles of isolated point sources in our
ACS/WFC science images), adjusted by aperture corrections calculated
on the basis of {\sc TinyTim} model PSFs. We adopted sky annuli with
inner and outer radii of 6 and 8 pixels, respectively, for local
background subtraction, a choice based on inspection of the radial
intensity distributions of a large subsample of our target
clusters. We generated artificial PSFs for each filter/camera
combination to determine the specific aperture corrections
required. (Note that this approach implicitly assumes that all of our
sample clusters are genuine point sources.) The resulting photometric
database is included in Table \ref{photom.tab}. Throughout this paper
and in all relevant figures, we have adopted the {\sc STmag} system,
whose photometric zero points (zpt) are defined by the {\sl HST} image
header keywords {\sc photflam} (see Table \ref{data.tab}) and {\sc
  photzpt} = $-21.1$, through zpt = $-2.5$ log({\sc photflam}) + {\sc
  photzpt}.

The aperture corrections required for 3-pixel source radii and
6--8-pixel sky annuli (for 0.05 arcsec pixels) are $-0.18, -0.19,
-0.19, -0.21, -0.46$ and $-0.63$ mag, respectively, for the F330W,
F547M, F606W, F814W, F110W and F160W photometry. The uncertainties
associated with our photometry, determined using standard
aperture-photometry routines implemented in our IDL package, are owing
to the necessarily limited number of photons detected (i.e., they
represent the photon noise in the background annuli owing to the
discrete nature of the incoming photons) and to errors in determining
the sky region's contribution (particularly in the case of
non-constant background levels). In essence, therefore, the
uncertainties reflect the signal-to-noise ratios of and fluctuations
in the background annuli. The flux levels in our background annuli
correspond to the {\it median} fluxes, which is an adequate
representation of the background contribution even in highly variable
regions such as circumnuclear starburst rings.  Using the detectors'
`gain' parameters listed in the image headers, the standard aperture
photometry routine determined the uncertainties in units of the
photo-electrons triggered by incident photons, while also taking into
account the detectors' read-out noise characteristics, also provided
as image header parameters.

\begin{table*}
 \begin{center}
  \caption{Aperture-corrected medium- and broad-band photometry of the
    104 NGC 6951 cluster candidates using fixed apertures$^a$}
  \label{photom.tab}
{\tiny
  \begin{tabular}{@{}rccccccccccccccccc@{}}
  \hline
\# & $\Delta \alpha$ & $\Delta \delta$ & $m_{\rm F330W}$ & $m_{\rm F547M}$ & $m_{\rm F606W\
}$ & $m_{\rm F814W}$ & $m_{\rm F110W}$ & $m_{\rm F160W}$ & XID$^b$\\
& (arcsec) & (arcsec) & (mag) & (mag) & (mag) & (mag) & (mag) & (mag) \\
 \hline
1 & $-$0.07 & $-$4.41 & $23.93 \pm 1.17$ & $22.81 \pm 0.49$ & $22.71 \pm 0.24$ & $22.60 \pm 0.13$ & $20.78 \pm 0.15$ & $20.50 \pm 0.14$ & $\cdots$ \\
2 & $-$7.82 & $-$4.15 & $23.01 \pm 0.83$ & $20.75 \pm 0.19$ & $20.51 \pm 0.08$ & $20.08 \pm 0.04$ & $18.77 \pm 0.05$ & $18.52 \pm 0.05$ & $\cdots$ \\
3 &    1.43 & $-$4.03 & $23.35 \pm 0.98$ & $23.18 \pm 0.79$ & $23.50 \pm 0.83$ & $23.85 \pm 1.04$ & $23.36 \pm 2.79$ & $24.67 \pm 0.99$ & $\cdots$ \\
4 &    0.15 & $-$3.84 & $21.94 \pm 0.48$ & $21.83 \pm 0.35$ & $21.17 \pm 0.13$ & $20.91 \pm 0.08$ & $19.65 \pm 0.13$ & $19.21 \pm 0.11$ & 53 \\
5 &    2.28 & $-$3.68 & $21.71 \pm 0.55$ & $22.41 \pm 0.66$ & $22.28 \pm 0.32$ & $22.91 \pm 0.50$ & $21.72 \pm 0.56$ & $21.37 \pm 0.50$ & 52\\
$\cdots$ & $\cdots$ & $\cdots$ & $\cdots$ & $\cdots$ & $\cdots$ & $\cdots$ & $\cdots$& $\cdots$ & $\cdots$ \\
\hline
\end{tabular}
}
\end{center}
\flushleft
$^a$ Coordinates are given with respect to the galaxy's centre,
 R.A. (J2000) = 20$^{\rm h}$ 37$^{\rm m}$ 14.13$^{\rm s}$, Dec (J2000)
 = 66$^{\circ}$ 06$'$ 20.2$''$.\\
$^b$ Cross-identification: ID number from vdL13.\\ 
The photometric uncertainties adopted in this paper reflect the
signal-to-noise ratios of and fluctuations in the background annuli
(see the text).\\
Table \ref{photom.tab} is published in its entirety in the electronic
edition of the journal, as supplementary data file. A portion is shown
here for guidance regarding its form and content.
\end{table*}

\begin{table*}
 \begin{center}
  \caption{Variable-aperture medium- and broad-band photometry of the
    104 NGC 6951 cluster candidates}
  \label{photom2.tab}
{\tiny
  \begin{tabular}{@{}rcccccccccccccc@{}}
  \hline
\# & $m_{\rm F330W}$ & $m_{\rm F547M}$ & $m_{\rm F606W\
}$ & $m_{\rm F814W}$ & $m_{\rm F110W}$ & $m_{\rm F160W}$ \\
& (mag) & (mag) & (mag) & (mag) & (mag) & (mag) \\
 \hline
1 & $23.31 \pm 0.90$ & $21.66 \pm 0.30$ & $21.50 \pm 0.16$ & $21.46 \pm 0.10$ & $\cdots$         & $19.08 \pm 0.09$ \\
2 & $23.67 \pm 1.08$ & $20.75 \pm 0.19$ & $20.49 \pm 0.08$ & $20.04 \pm 0.03$ & $18.74 \pm 0.05$ & $18.65 \pm 0.05$ \\
3 & $22.43 \pm 0.63$ & $21.99 \pm 0.39$ & $21.99 \pm 0.28$ & $21.82 \pm 0.14$ & $22.28 \pm 1.09$ & $22.79 \pm 2.49$ \\
4 & $20.79 \pm 0.28$ & $20.62 \pm 0.21$ & $20.60 \pm 0.15$ & $20.03 \pm 0.04$ & $18.63 \pm 0.10$ & $18.33 \pm 0.10$ \\
5 & $22.50 \pm 0.63$ & $22.45 \pm 0.46$ & $22.47 \pm 0.30$ & $22.68 \pm 0.26$ & $21.26 \pm 0.28$ & $21.13 \pm 0.34$ \\
$\cdots$ & $\cdots$ & $\cdots$ & $\cdots$ & $\cdots$ & $\cdots$& $\cdots$ \\
\hline
\end{tabular}
}
\end{center}
\flushleft
Table \ref{photom2.tab} is published in its entirety in the electronic
edition of the journal, as supplementary data file. A portion is shown
here for guidance regarding its form and content.
\end{table*}

Small-aperture photometry combined with aperture corrections is
applied increasingly often (e.g., Harris et al. 2004; Annibali et
al. 2011; vdL13), even for extended objects. However, while this
approach is well-established for point-source photometry, it breaks
down for extended objects (see, e.g., de Grijs et al. 2013a), unless
one fully models the radial profiles (Anders et al. 2006). To show
this, we additionally adopted an adaptive aperture-photometry approach
(de Grijs \& Anders 2012; de Grijs et al. 2013a; Li et al. 2015),
scaling the source apertures and sky annuli based on a measure of the
objects' sizes (i.e., their Gaussian sizes, $\sigma_{\rm G}$). We used
a source aperture radius of $3 \sigma_{\rm G}$, and $3.5\sigma_{\rm
  G}$ and $5\sigma_{\rm G}$ for the inner and outer sky annuli,
respectively. The exact scaling was determined by checking the stellar
growth curves, to identify where the objects' radial profiles vanish
into the background noise. We have confirmed that the choice of our
source radii was conservative enough so as not to miss any genuine
source flux, and we also verified that our selected sky annuli were
not significantly contaminated by neighbouring sources. We adjusted
the source apertures and sky annuli where necessary to minimize the
effects of crowding. The resulting photometric database is included in
Table \ref{photom2.tab}. As before, the uncertainties associated with
our object photometry reflect the signal-to-noise ratios of and
fluctuations in the background annuli.

\begin{figure}
\begin{center}
\includegraphics[width=\columnwidth]{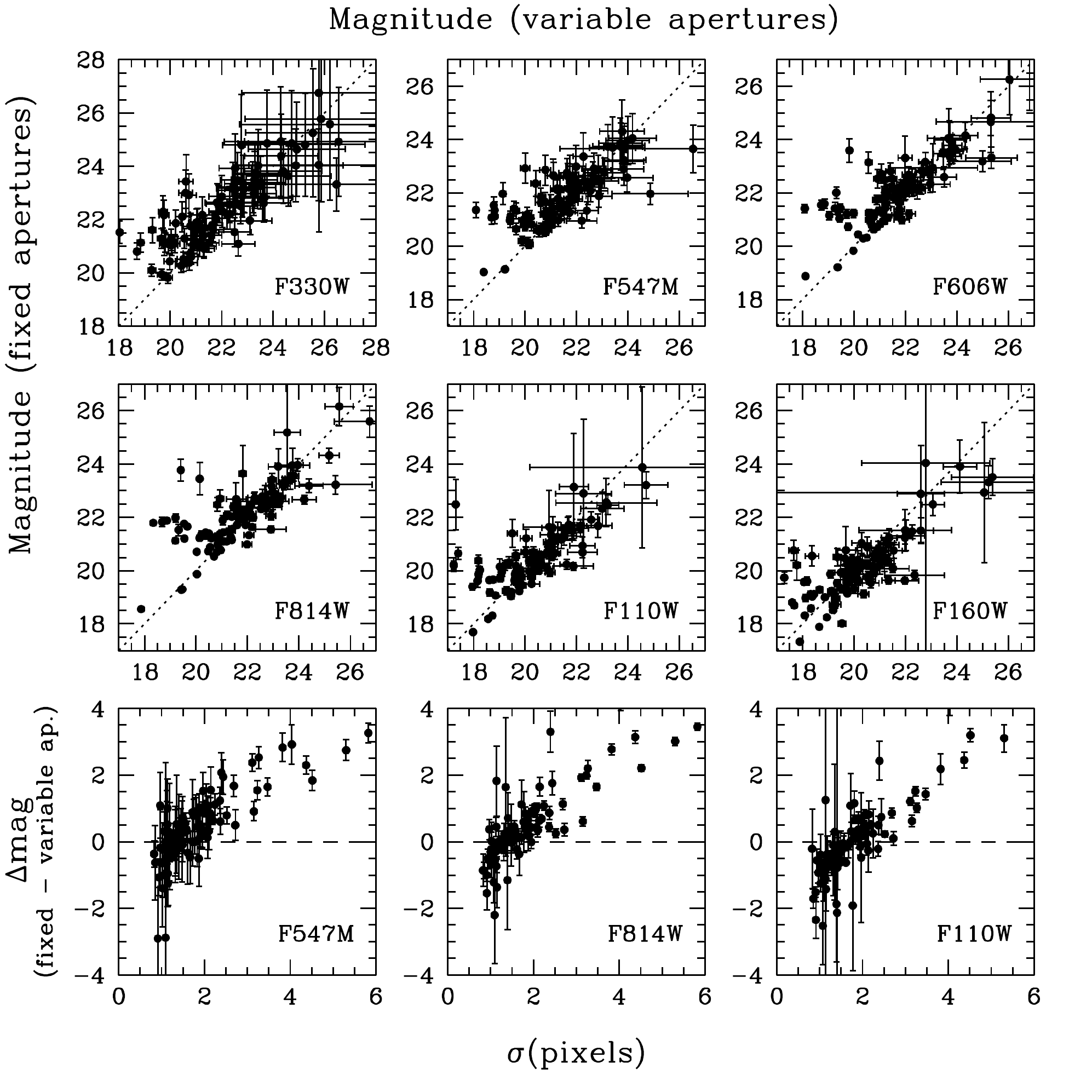}
\caption{Comparison of our aperture-corrected PSF photometry for the
  NGC 6951 cluster population with that resulting from the adaptive
  aperture-photometry approach. Top and middle rows: individual
  comparisons for all six medium- and broad-band filters used. Refer
  to the top axis label. Bottom row: magnitude differences (fixed
  minus variable source apertures) as a function of characteristic
  Gaussian object size for three representative filters.}
\label{cfphotom.fig}
\end{center}
\end{figure}

Figure \ref{cfphotom.fig} offers a comparison of our
aperture-corrected PSF photometry with that resulting from the
adaptive aperture-photometry approach in all medium- and broad-band
filters used here. The panels in the top and middle rows show the
detailed comparisons. It is clear that while a significant subset of
our sample clusters are well matched using either approach, a large
minority of sources are fainter in the fixed-aperture photometry. The
panels in the bottom row of Fig. \ref{cfphotom.fig} show that these
objects tend to be significantly more extended than the canonical PSF
size. We visually checked these objects' radial profiles in relation
to the aperture sizes adopted. We are confident that our choice of
variable aperture size has resulted in inclusion of all of the
objects' photometry. Indeed, this result underscores our reluctance to
rely on aperture-corrected PSF photometry since that approach ignores
the radial intensity distributions of more extended sources and,
hence, often causes one to underestimate the relevant source fluxes.

\begin{figure}
\begin{center}
\includegraphics[width=\columnwidth]{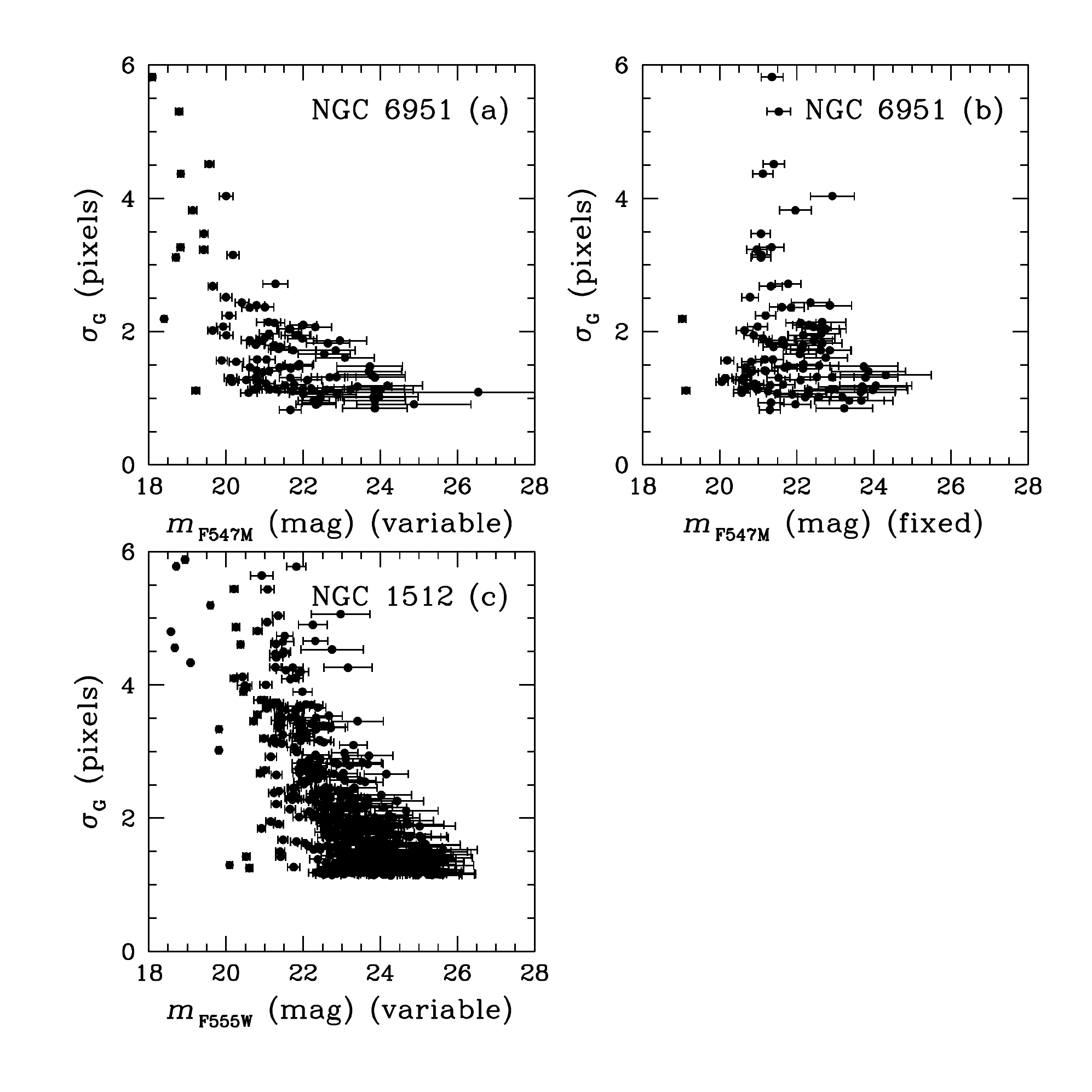}
\caption{NGC 6951 cluster magnitudes in the F547M filter versus size
  ($\sigma_{\rm G}$, in units of ACS/WFC pixels) for (a) variable and
  (b) fixed source apertures. (c) NGC 1512 $m_{\rm F555W}$ photometry
  as a function of size (WFC3 pixels) for variable source apertures.}
\label{magsize.fig}
\end{center}
\end{figure}

In Fig. \ref{magsize.fig}a,b, we show the NGC 6951 clusters' Gaussian
sizes, $\sigma_{\rm G}$, as a function of their F547M magnitudes. It
is clear that while the brighter objects tend to exhibit larger
sizes,\footnote{If---as expected for well-sampled clusters---light
  traces mass, this is a direct consequence of the mass--size
  relation.} the fainter objects are significantly more compact. We
attempted to excise star cluster {\it complexes} from our final sample
on the basis of the objects' profile fits. The size of the ACS/WFC PSF
is $\sigma_{\rm G,PSF} \simeq 0.9 \mbox{ pix} \simeq 5$ pc at the
distance of NGC 6951. Of order half a dozen objects with $\sigma_{\rm
  G}$'s often well in excess of 10 pixels ($\sigma_{\rm G} \ga 55$ pc)
were discarded. The detection of a small number of objects with very
large sizes was mostly caused by crowding in the images; we carefully
checked that our final sample objects are not severely affected by
crowding. Figure \ref{magsize.fig} thus shows that we have been
successful at distinguishing genuine objects from spurious detections.

\begin{figure}
\begin{center}
\includegraphics[width=\columnwidth]{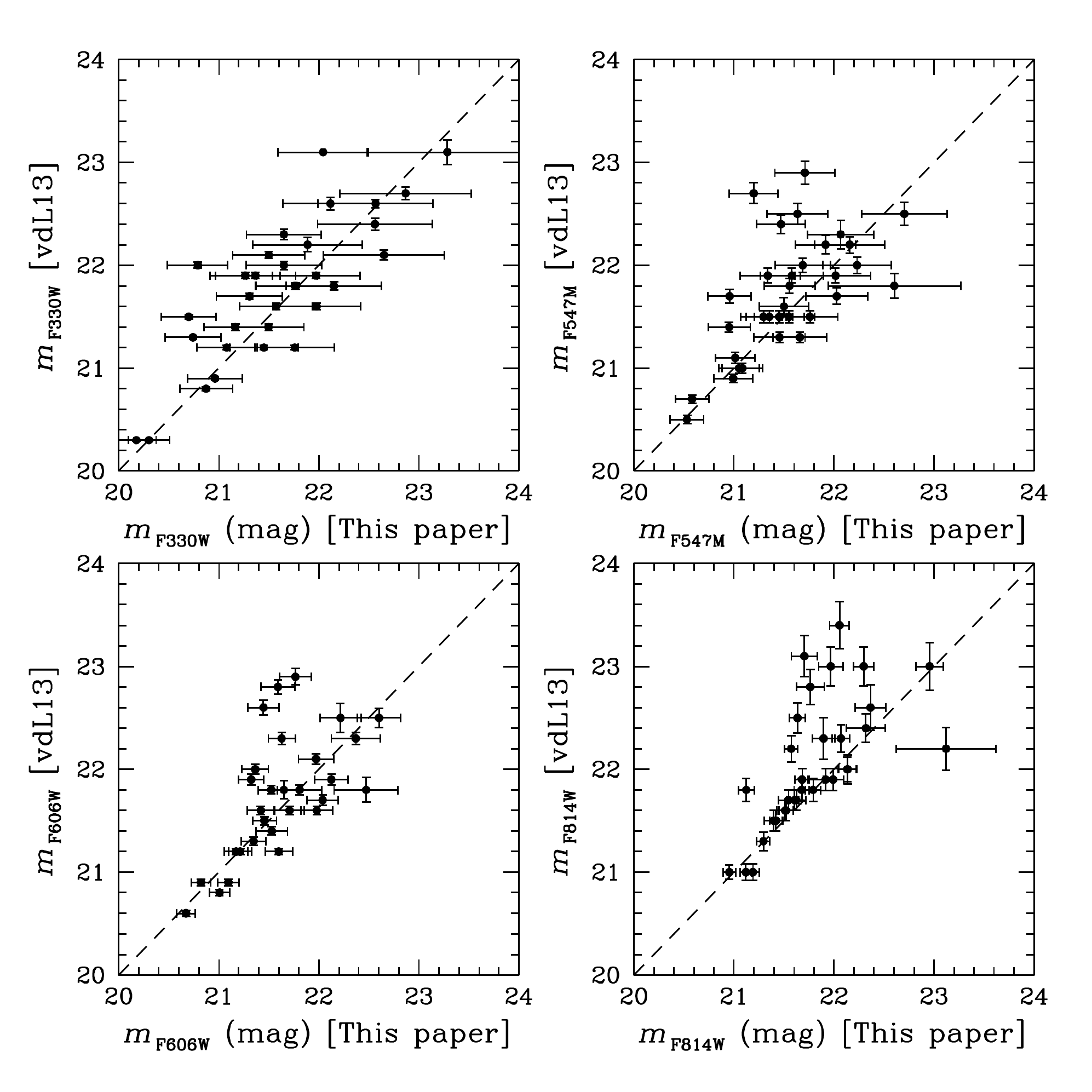}
\caption{Comparison of our aperture-corrected NGC 6951 PSF cluster
  photometry with that of vdL13 in the F330W, F547M, F606W and F814W
  filters.}
\label{cfphotom2.fig}
\end{center}
\end{figure}

Figure \ref{cfphotom2.fig} shows a comparison of our
aperture-corrected PSF photometry with that of vdL13 in the F330W,
F547M, F606W and F814W filters. Of the 55 objects identified by vdL13,
we found 38 unambiguous matches in our database; an additional seven
objects could not be uniquely cross-identified, either because one of
our objects appeared to coincide with two objects in the vdL13
database (three cases) or vice versa. In addition, vdL13's objects 3,
9, 18, 20, 21, 24 and 35 do not have counterparts in our
database. vdL13 obtained their photometry based on a combination of
{\sc SExtractor} and {\sc daophot} routines. Specifically, they
decided to use source apertures of 2 pixels (0.09 arcsec) and sky
annuli between 6 and 16 pixels, in essence to avoid the effects of
crowding. On the whole -- and within the observational uncertainties
and given the cluster size-dependent systematics introduced by the use
of fixed apertures combined with aperture corrections
(Fig. \ref{cfphotom.fig}) -- our PSF photometry is indeed consistent
with the vdL13 fixed-aperture photometry, confirming our analysis in
Fig. \ref{cfphotom.fig}.

Next, we obtained photometric measurements for our 469 NGC 1512 sample
clusters in the targeted central galaxy field (in both the
circumnuclear starburst ring and the main galactic disk region) to
acquire their SEDs using our adaptive aperture-photometry
method. Table \ref{photom3.tab} includes the photometric measurements
for the 275 sample clusters with good-quality measurements, which we
defined as objects with observational uncertainties $\le 1.0$ mag in
at least three filters. Figure \ref{magsize.fig}c shows the clusters'
Gaussian sizes (expressed in units of pixels) as a function of their
F555W magnitudes. The trend seen for the NGC 1512 clusters mimicks
that seen for the NGC 6951 cluster sample.

\begin{table*}
\begin{center}
\begin{minipage}{130mm}
\caption{Photometry of the 275 NGC 1512 cluster candidates with good-quality measurements\label{photom3.tab}}
{\tiny
\begin{tabular}{ccccccccc}
\hline
\#&\multicolumn{2}{c}{R.A. (J2000)} & \multicolumn{2}{c}{Dec. (J2000)}&$m_{\rm F336W}$&$m_{\rm F438W}$&$m_{\rm F555W}$&$m_{\rm F814W}$\\
 &$(^{\circ})$&(hh mm ss.ss)&$(^{\circ})$& $(^{\circ} \, ' \, '')$ & (mag) & (mag) & (mag) & (mag)\\
\hline
1&60.97376&04 03 53.70&$-$43.34726&$-$43 20 50.14&$22.58 \pm 0.56$&$21.65 \pm 0.26$&$21.38 \pm 0.13$&$22.09 \pm 0.17$\\
2&60.97379&04 03 53.71&$-$43.34728&$-$43 20 50.22&$22.72 \pm 0.60$&$22.23 \pm 0.34$&$22.44 \pm 0.22$&$23.14 \pm 0.29$\\
3&60.97404&04 03 53.76&$-$43.34718&$-$43 20 49.85&$22.17 \pm 0.46$&$21.65 \pm 0.26$&$21.44 \pm 0.12$&$21.72 \pm 0.13$\\
4&60.97411&04 03 53.78&$-$43.34715&$-$43 20 49.76&$24.70 \pm 1.51$&$23.05 \pm 0.51$&$22.91 \pm 0.24$&$23.31 \pm 0.26$\\
5&60.97289&04 03 53.49&$-$43.34803&$-$43 20 52.92&$23.45 \pm 0.831$&$23.162 \pm 0.527$&$23.085 \pm 0.268$&$23.97 \pm 0.376$\\
$\cdots$ & $\cdots$ & $\cdots$ & $\cdots$ & $\cdots$ & $\cdots$ & $\cdots$ & $\cdots$ & $\cdots$ \\
\hline
\end{tabular}
}
\end{minipage}
\end{center}
\flushleft
Table 2 is published in its entirety in the electronic version of the
journal, as a supplementary file. A portion is shown here for guidance
regarding its form and content.
\end{table*}

\subsection{Completeness checks}
\label{compl.sec}

We next proceeded to characterize the observational completeness
levels. For our completeness tests of the NGC 6951 data, we selected
the area of $350 \times 350$ pixels$^2$ centred on the galaxy's centre
shown in Fig. \ref{newfig1} (top). This region is dominated by the
circumnuclear starburst ring ($35 \la R \la 105$ pixels). For a given
filter, we first added 500 artificial extended sources (all assigned
the same magnitude and of the same size, $\sigma_{\rm G} = 2$ pixels)
with random ($X$, $Y$) coordinates to our science frames, carefully
avoiding positions too close to the science images' edges. The
positions of 163 of these artificial sources coincided with the area
occupied by the starburst ring. Note that the majority of our sample
clusters have sizes somewhat smaller than the size adopted for our
artificial sources (see Fig. \ref{magsize.fig}). This implies that the
resulting completeness levels are conservative (everything else being
equal, these completeness limits are somewhat brighter than expected
for the `average' object). The completeness levels for objects with a
smaller Gaussian (point-source-like) size of $\sigma_{\rm G} = 1$
pixel are $\sim$0.4--0.7 mag fainter from the bluest to the reddest
filters in our collection.

\begin{figure}
\begin{center}
\includegraphics[width=\columnwidth]{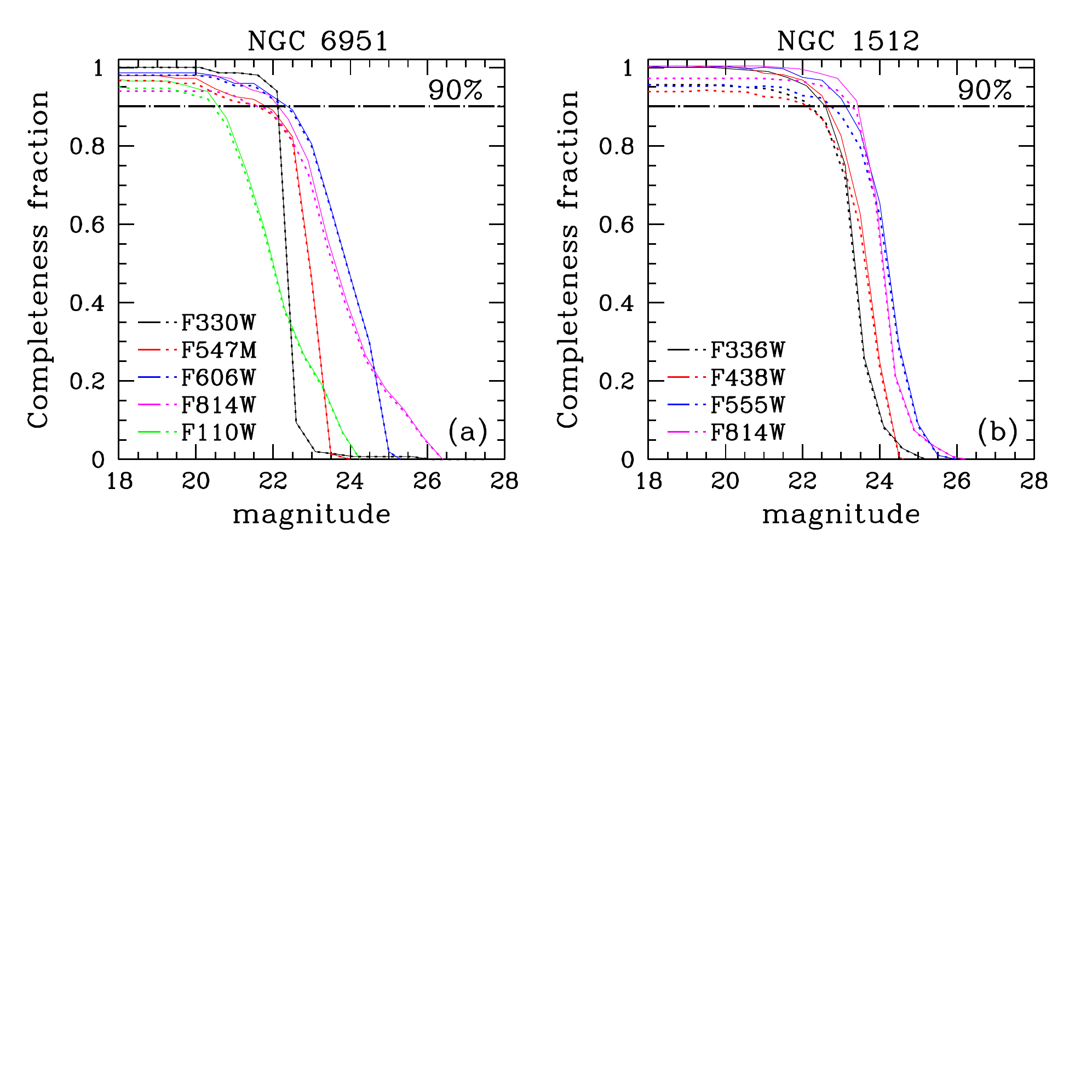}
\vspace{-4cm}
\caption{(a) Completeness fractions for $\sigma_{\rm G} = 2$ pixel
  source detection in the NGC 6951 ring region in the main filters
  used in this paper. The 90 per cent completeness limit is
  highlighted by means of the horizontal dash-dotted line. Magnitudes
  are given in the {\sc STmag} system. (b) As in panel (a) but for NGC
  1512 and $\sigma_{\rm G} = 2.5$ pixels.}
\label{fig2}
\end{center}
\end{figure}

In retrieving the input artificial clusters, we first checked whether
we could detect them in both the science and template images in both
filters (which helped us to evaluate how many artificial sources could
be recovered after correction for blends and chance superpositions),
using the same source discovery routine as applied to detect the real
objects. Of the 500 input sources, 476 could be retrieved from images
containing only artificial sources and no background flux prior to any
completeness analysis; 4--9 blends with real objects were detected,
depending on the filter considered, all of which were found in the
region occupied by the ring. We subsequently applied our
cross-identification technique to the artificial cluster lists in both
middle-wavelength filters used for the object selection to further
consolidate their authenticity. This step led to rejection of those
clusters that were too faint in one of the input filters to meet our
minimum detection limit. We subsequently obtained photometry for the
resulting cluster catalogue. Figure \ref{fig2} displays the
completeness levels of our science frames as a function of input
magnitude for all filters used. We additionally ran completeness tests
for the ring region only, following exactly the same approach as for
the full frames.

In the F330W, F547M, F606W, F814W and F110W filters, the 90 per cent
completeness limits are $m_{\rm F330W}^{90\%} = 22.1$ mag, $m_{\rm
  F547M}^{90\%} = 21.8$ mag, $m_{\rm F606W}^{90\%} = 22.4$ mag,
$m_{\rm F814W}^{90\%} = 22.1$ mag and $m_{\rm F110W}^{90\%} = 20.5$
mag. Note that the 90 per cent completeness limits occur at $\sim
0.5$--1.0 mag brighter magnitudes than the conventionally used 50 per
cent completeness limits, in all filters. We point out that our
selection limit for clusters in NGC 6951 is ultimately determined by
the completeness levels in the F547M filter, the filter we used for
our source detection characterized by the lowest completeness. This is
driven by our data reduction procedures: we first determined the
positions of all genuine objects and subsequently obtained the
photometry at those positions in all filters, irrespective of
completeness level. The resulting photometric databases (Tables
\ref{photom.tab} and \ref{photom2.tab}) show that this results in
reliable photometric determinations in all filters to well below the
canonical 90 per cent levels for any of the other filters (i.e., other
than F547M). Finally, the 90 per cent completeness level in the F547M
filter corresponds to an absolute magnitude of $M_V \sim -10.1$ mag,
which underscores that our sample consists of genuine star clusters
only: individual blue supergiants do not reach these brightnesses
(e.g., Maoz et al. 2001).

To obtain the corresponding completeness levels pertaining to our
central NGC 1512 images (covering $585 \times 585$ pixels$^2$), we
proceeded similarly as for NGC 6951, using 306 artificial sources as
completeness benchmark in each filter and for each step in
magnitude. The artificial cluster size adopted was based on the size
distribution of our genuine cluster sample, which exhibits a
significant peak at sizes $\sigma_{\rm G} \le 2.5$ pixels (see, e.g.,
Fig. \ref{magsize.fig}, which also shows that the larger sources tend
to be brighter and, hence, will not affect our completeness analysis
significantly): we thus assigned the median $\sigma_{\rm G} = 2.2$
pixels ($\equiv 5.3$ pc) to our artificial clusters. Of the 306 input
sources, 294 could be retrieved from images containing only artificial
sources and no background flux prior to any completeness analysis;
8--11 blends with real objects were detected, depending on the filter
considered, all of which were found in the region occupied by the
ring.

The resulting completeness levels of our science frames are shown in
Fig. \ref{fig2} as a function of input magnitude, for all filters. We
found that for the F336W, F438W, F555W and F814W filters, the
corresponding 90 per cent completeness limits are $m_{\rm
  F336W}^{90\%} = 22.6$ mag, $m_{\rm F438W}^{90\%} = 22.7$ mag,
$m_{\rm F555W}^{90\%} = 23.2$ mag and $m_{\rm F814W}^{90\%} =23.5$
mag. The more commonly used 50 per cent completeness limits occur at
brightness levels that are fainter by 0.6 mag, 1.0 mag, 1.0 mag and
0.7 mag, respectively. From Fig. \ref{fig2}, we can see that the F336W
filter is our limiting filter for the NGC 1512 analysis. However,
since our object selection is based on source detection and
cross-identification in the F438W and F555W filters, our sample is
intrinsically limited by the least sensitive of the latter
filters. Our NGC 1512 cluster sample is therefore a `$B$-band
detection limited' sample. Its 90 per cent completeness level
corresponds to an absolute magnitude of $M_{\rm F438W}^{90\%} = -7.3$
mag. Although individual blue supergiants are sufficiently luminous to
be seen at these brightness levels, their numbers are expected to be
too small to affect the statistical results obtained in this paper
(e.g., Maoz et al. 2001; Chandar et al. 2010).

\section{Cluster luminosity functions}
\label{clfs.sec}

The CLFs are among the most important diagnostics considered in
discussing the evolution of statistically well-sampled star cluster
populations in external galaxies. It is well-established that shortly
after the time of cluster formation, both the CLF and its
corresponding CMF are expected to resemble power-law distributions,
i.e., $N(L) \propto L^{-\alpha_1}$ and $N(M) \propto M^{-\alpha_2}$,
where $\alpha_1$ and $\alpha_2$ usually attain values close to 2
(cf. de Grijs et al. 2003b; Portegies Zwart et al. 2010; Fall \&
Chandar 2012). As discussed in Section 1, over time these populations
lose their lower-mass and, hence, lower-luminosity members, thus
resulting in rounded distributions that are often approximated by a
lognormal distribution (e.g., Fall \& Zhang 2001; Fall 2006). In this
evolutionary context, it is important to define CLFs (and CMFs) that
cover only narrow age ranges in order to avoid complications owing to
poorly quantified evolutionary effects.

\begin{figure}
\begin{center}
\includegraphics[width=\columnwidth]{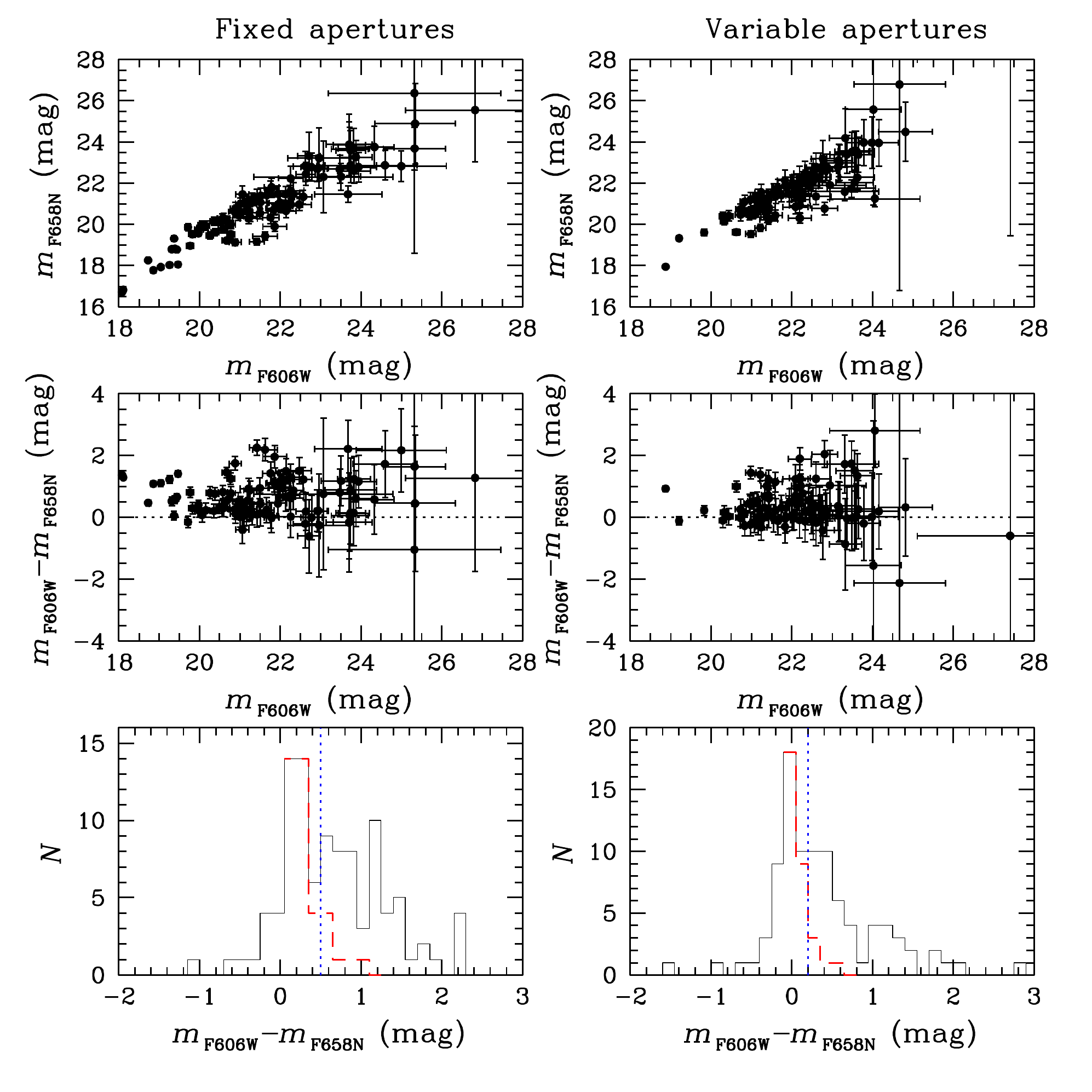}
\caption{Top and middle rows: Comparison of the NGC 6951 H$\alpha$
  emission-line (F658N) versus continuum (F606W) cluster photometry
  (see text). Bottom row: Distributions of the differences in
  magnitude between the F606W and F658N fluxes. Red dashed lines:
  Mirrored histograms containing objects without H$\alpha$
  emission-line excess flux. Blue dotted lines: Minimum magnitude
  differences adopted for selection of our H$\alpha$-excess samples.}
\label{halphacf.fig}
\end{center}
\end{figure}

We are fortunate in also having access to {\sl HST} observations of
both NGC 6951 and NGC 1512 which were obtained through narrow-band
filters centred on the H$\alpha$ emission line. We obtained ACS/WFC
observations of the NGC 6951 starburst ring from the HLA which were
observed through the F658N filter. The H$\alpha$ emission line, at a
rest-frame wavelength of $\lambda = 656.28$ nm, appears only for very
young stellar populations, with ages up to approximately 10
Myr. Selection of a subsample of our target clusters based on excess
H$\alpha$ emission (with respect to the flux level expected from the
continuum; see below) will thus ensure that we only include the
youngest clusters. We thus proceeded to obtain our sample clusters'
H$\alpha$ photometry adopting exactly the same source positions and
aperture radii as used for the medium- and broad-band filters in our
observational arsenal, followed by standard flux calibration using the
{\sc photflam} {\sl HST} image header keyword (see Table
\ref{data.tab}, top).

Figure \ref{halphacf.fig} forms the basis of our selection of
H$\alpha$-excess objects. We provide the results for both (left)
aperture-corrected fixed-aperture and (right) size-dependent
variable-aperture photometry. The panels in the top and middle rows
show direct comparisons of the emission-line (F658N) and continuum
(F606W) fluxes of our sample sources. The panels in the bottom row
show the distributions of the differences in magnitude between the
F606W and F658N fluxes. Objects without excess H$\alpha$ emission
(i.e., clusters older than approximately 10 Myr) are expected to
reside in the histograms' peaks centred on a zero magnitude
difference. In order to allow us to select H$\alpha$-excess objects,
we mirrored the left-hand sides of the histograms, as shown by the red
dashed lines. The blue dotted lines show the minimum magnitude
differences adopted for selection of H$\alpha$-excess objects, i.e.,
$(m_{\rm F606W} - m_{\rm F658N}) \ge 0.5$ mag for our fixed-aperture
photometry and $(m_{\rm F606W} - m_{\rm F658N}) \ge 0.2$ mag for our
variable apertures. For NGC 6951, these choices led to
`H$\alpha$-excess samples' containing 55 and 49 objects,
respectively. In the following, we will also show the results for
additional, more strictly defined H$\alpha$-excess samples
characterized by $(m_{\rm F606W} - m_{\rm F658N}) \ge 1.0$ mag (for
both sets of photometric measurements) in order to underscore the
reliability of our results. The latter samples are, however, more
significantly affected by small-number statistics.

\begin{figure}
\begin{center}
\includegraphics[width=\columnwidth]{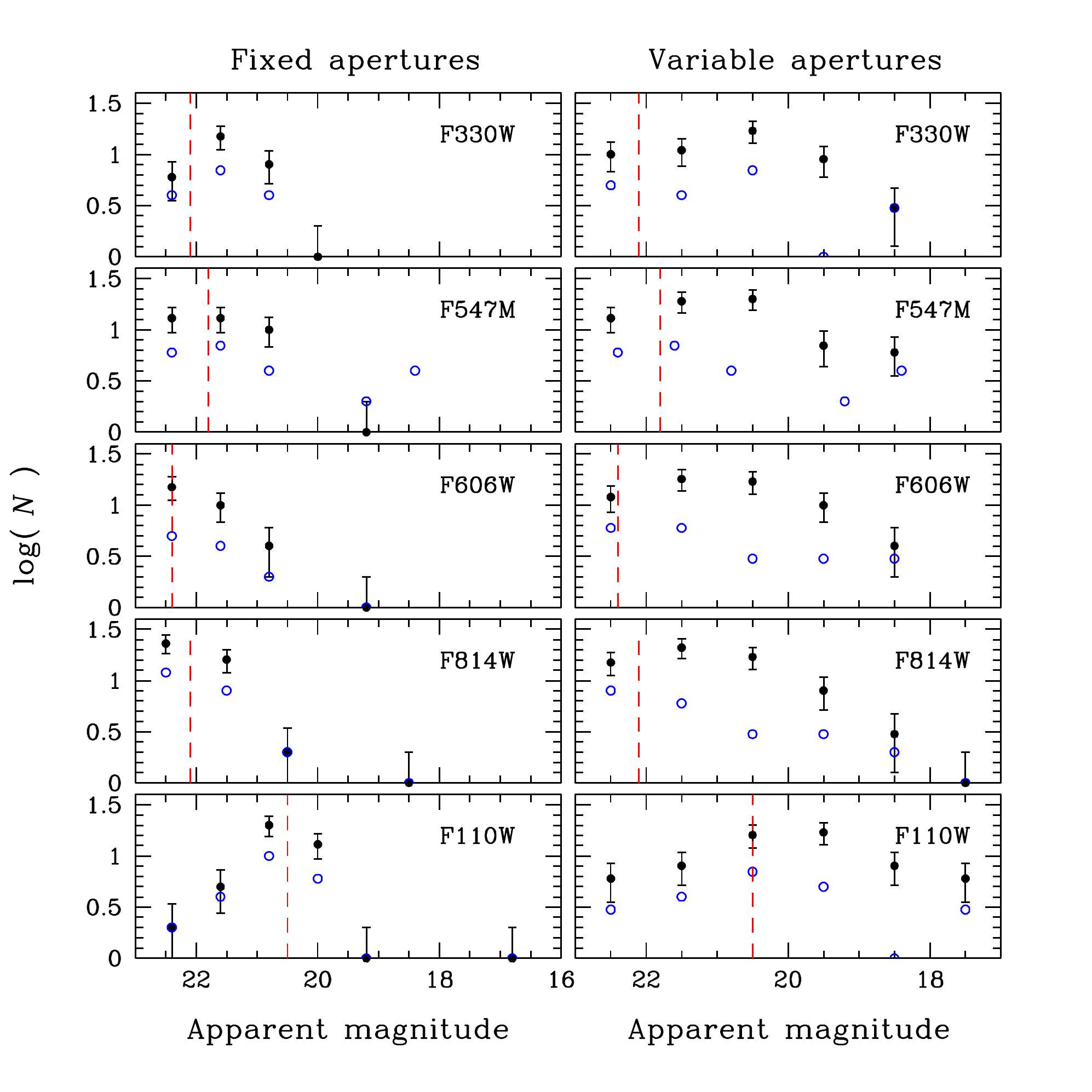}
\caption{CLFs composed of the young, H$\alpha$-excess objects in NGC
  6951. The red vertical dashed lines represent the 90 per cent
  completeness limits in the ring area in the relevant filters; error
  bars denote Poissonian uncertainties. Black solid bullets: CLFs for
  $(m_{\rm F606W} - m_{\rm F658N}) \ge 0.5$ mag (fixed apertures) and
  $(m_{\rm F606W} - m_{\rm F658N}) \ge 0.2$ mag (variable
  apertures). Blue open circles: CLFs for $(m_{\rm F606W} - m_{\rm
    F658N}) \ge 1.0$ mag (both fixed and variable apertures).}
\label{n6951lfs.fig}
\end{center}
\end{figure}

\begin{figure}
\begin{center}
\includegraphics[width=1.2\columnwidth]{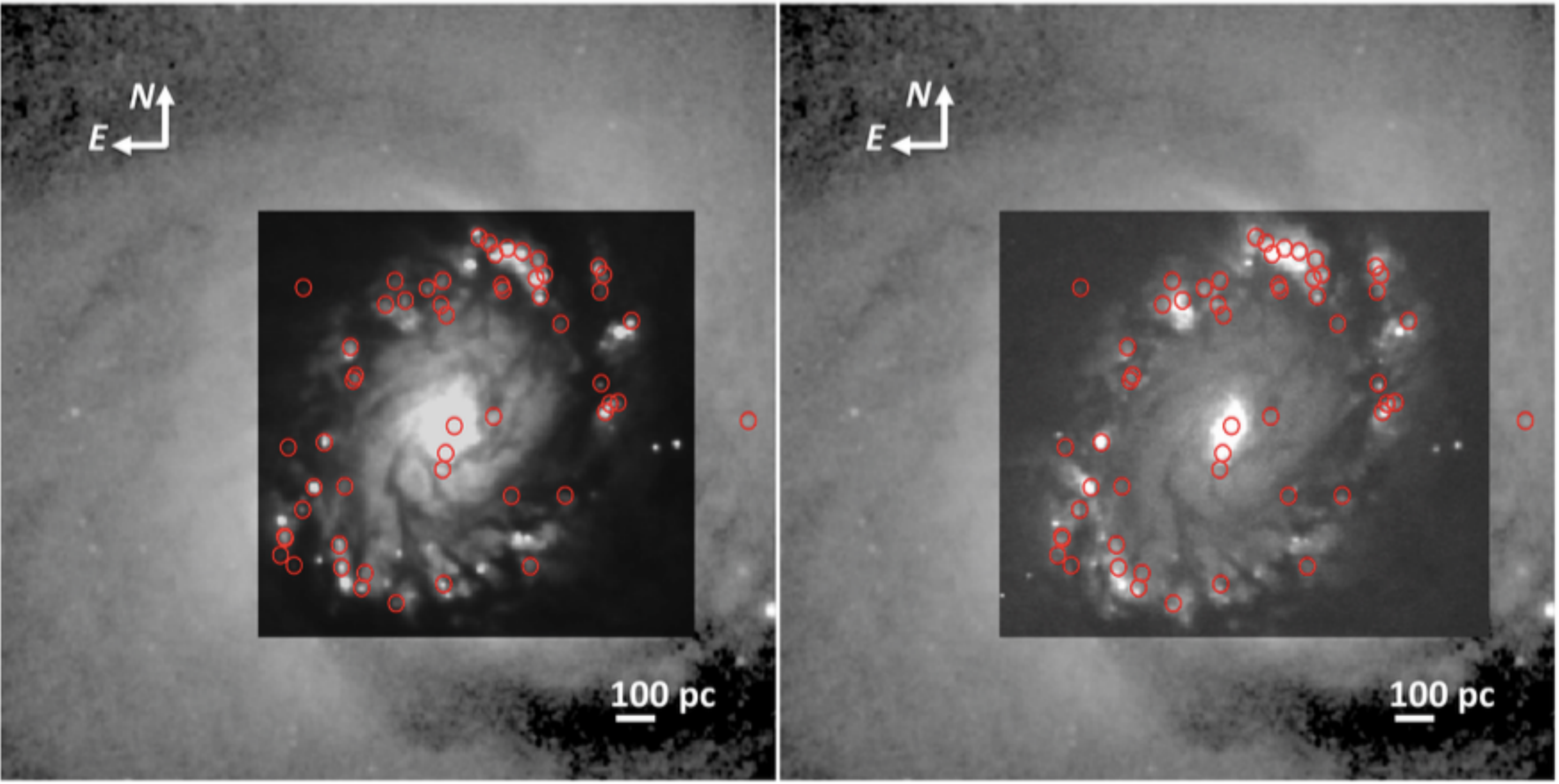}
\caption{As Fig. \ref{newfig1} (top) but showing the H$\alpha$-excess
  objects overplotted on (left) the continuum F606W image and (right)
  the H$\alpha$ (F658N) line image.}
\label{hafig}
\end{center}
\end{figure}

Figure \ref{hafig} shows the H$\alpha$-excess objects in NGC 6951
overplotted on both a continuum image observed through the F606W
filter and the H$\alpha$ (F658N) image. The resulting CLFs for the
young, H$\alpha$-excess objects in NGC 6951 are shown in
Fig. \ref{n6951lfs.fig}. We have also indicated the completeness
levels in all panels. Note that, strictly speaking, our samples are
F547M-limited samples; the completeness limits for the other filters
in our arsenal are for illustrative purposes only. The youngest
clusters, characterized by an H$\alpha$ excess, are found both in a
broad swath due north of the galactic centre and in a concentration in
the southwest of the starburst ring. vdL13 analysed the age
distribution of their 55 sample clusters in the NGC 6951 starburst
ring and tentatively suggested that the clusters may be younger in
both the southwestern and northeastern quadrants. Careful comparison
with their data tables, as well as inspection of their Fig. 7, shows
that our results are in broad agreement with theirs.

We proceeded similarly for NGC 1512. We compared the galaxy's cluster
photometry in the F555W broad-band filter with that in the adjacent
F658N narrow-band filter. Again, objects without excess H$\alpha$
emission are expected to reside in the histogram's peak centred on the
smallest $(m_{\rm F555W} - m_{\rm F658N})$ magnitude difference. (The
peak of the magnitude differences is slightly offset from zero
magnitude, because the F658N filter curve does not overlap with the
F555W filter curve, so that the slope of the SED becomes important.)
The minimum magnitude difference adopted for selection of
H$\alpha$-excess objects is $(m_{\rm F555W} - m_{\rm F658N}) \ge 1.0$
mag. We also imposed a brightness limit of $m_{\rm F555W} \ge 20.0$
mag. These choices led to `H$\alpha$-excess samples' containing 127
objects.

Before proceeding, we note one potential caveat which may have
affected our selection of `H$\alpha$-excess' clusters using their
F555W magnitudes. At low redshifts, observations of cluster
populations through this filter may be contaminated by the presence of
[OIII]4959,5007{\AA} line emission, which could be stronger than the
objects' H$\alpha$ emission (e.g., Reines et al. 2010, their fig. 11;
see also Anders \& Fritze--v. Alvensleben 2003). For very young
clusters, i.e., for ages below 3--4 Myr, the contribution of
[OIII]-emitting gas could be $\gtrsim 1.3$--1.5 mag brighter than that
from the stars alone, decreasing to a brightening of $\sim$1.0 mag at
4 Myr $\sim$0.3 mag at 6 Myr, and $\sim$0.15 mag at 8 Myr (Reines et
al. 2010). If our NGC 1512 cluster sample is indeed affected by
significant excess [OIII] flux in their F555W magnitudes, such
clusters may not show a significant deviation from the 1:1 relation
between their F658N and F555W magnitudes. If so, our
H$\alpha$-selected sample is conservative, implying that our method
may have missed a fraction of the youngest clusters. (We also note
that, at the youngest ages ($\lesssim$3--4 Myr), these clusters will
most likely still be embedded in their natal dust cocoons, so that
they may not have passed our selection criteria, which required
sufficient signal-to-noise ratios in the blue F438W images.)

\begin{figure}
\begin{center}
\includegraphics[width=1.1\columnwidth]{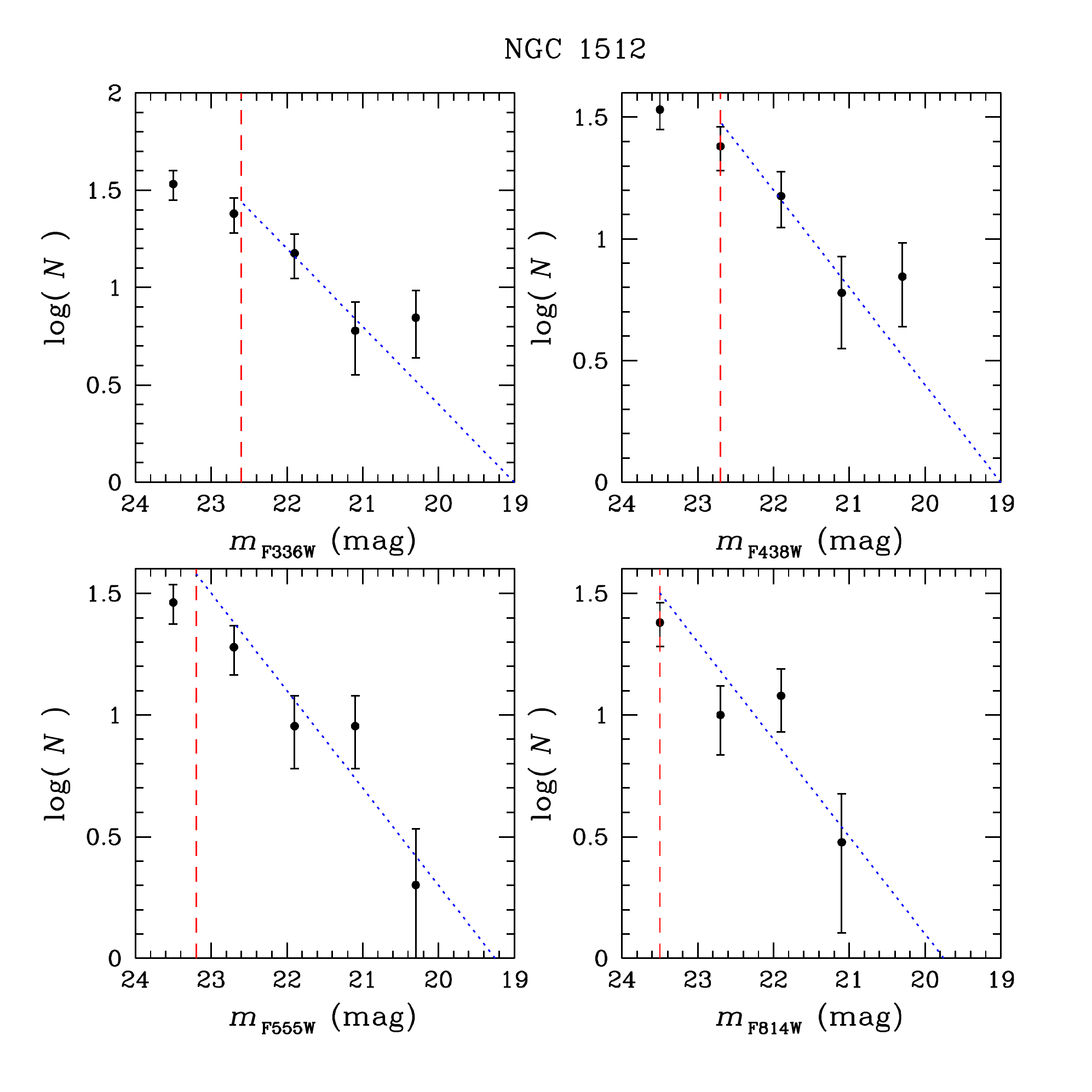}
\caption{CLFs of the H$\alpha$-selected NGC 1512 circumnuclear ring
  population in all passbands considered here, including their
  associated error bars (Poissonian uncertainties). The vertical red
  dashed lines are the 90 per cent completeness limits in the
  respective filters; the blue dotted lines indicate canonical $\alpha
  = 2$ power-law CLFs.}
\label{n1512lfs.fig}
\end{center}
\end{figure}

In Fig. \ref{n1512lfs.fig} we show the filter-dependent CLFs for the
cluster population detected in the NGC 1512 ring, including their
Poissonian uncertainties (shown as error bars). The red vertical solid
lines denote the respective 90 per cent completeness limits based on
our artificial-cluster tests. We also show the expected, canonical
$\alpha = 2$ power-law CLF slopes (blue dotted lines). Although the
CLFs are affected to some extent by small-number statistics at
brighter magnitudes, it is clear that -- within the observational
scatter -- the canonical slope holds well down to the 90 per cent
completeness limits for all passbands. In view of the caveat pointed
out above, it appears that contamination by [OIII] emission has not
significantly affected the CLF shapes of the H$\alpha$-selected
clusters compared with the full CLFs prior to any selection. Of
course, the impact of [OIII] contamination may also have affected our
sample selection equally across the observational magnitude range,
which is impossible to ascertain based on our current data
set. Nevertheless, if an intrinsic turnover in the youngest cluster
sample were somehow masked by excess [OII] emission, this would imply
a selection function in the F555W filter that appears to be at odd
with the selection functions in the other filters considered here.

The situation for NGC 6951 is more complex, however. As discussed
above, the galaxy's full CLFs in the F547M and F814W filters are well
represented by $\alpha = 2$ power laws which hold down to magnitudes
just above the 90 per cent completeness limits, where the CLFs appear
to turn over. Since they occur close to the detection limit, we do not
consider these apparent turnovers statistically significant. However,
the H$\alpha$-selected NGC 6951 CLFs shown in Fig. \ref{n6951lfs.fig}
start to turn over at significantly brighter magnitudes (by up to 1
mag) than the completeness limits in any of the filters
considered. The H$\alpha$-selected CLFs are composed of a subset of
the full CLFs, i.e., the former contain only those clusters that
exhibit clear H$\alpha$ emission and which are thus most likely
younger than approximately 10 Myr.

In a follow-up paper we will determine the clusters' ages and masses
based on their full SEDs, which will allow us to investigate this CLF
behaviour close to the completeness limits in more detail, and
specifically in terms of the CMF. Here, we suggest that the apparent
turnovers, occurring at up to 1 mag above the 90 per cent completeness
curves, are most likely caused by the H$\alpha$ magnitude cut. Fainter
young clusters, i.e., those located close to the completeness limits,
will also be fainter in terms of their H$\alpha$ emission. Given the
photometric uncertainties associated with the fainter sources,
H$\alpha$ selection will have predominantly removed fainter cluster
from the young subsample.

\section{Overall context}

We have presented a careful analysis of the CLFs of the starburst ring
clusters in NGC 6951 and NGC 1512. One of our main aims of the
analysis in this paper was to assess whether the dense and often
violent conditions in circumnuclear starburst rings may give rise to
differences in either the distribution of cluster luminosities at the
time of their formation, i.e., the initial CLF, or characteristic
disruption time-scale. To get an initial handle on answering this
question, we compiled a database of galaxies exhibiting circumnuclear
starburst rings and which were observed with the {\sl HST} in at least
four passbands. This latter criterion is necessary for successful SED
analysis to derive the clusters' age and mass distributions (e.g.,
Anders et al. 2004; de Grijs et al. 2013a). Among these, one of our
own earlier studies dealt with the starburst ring galaxy NGC 3310 (de
Grijs et al. 2003a,b).

Maoz et al. (2001) used {\sl HST}/WFPC2 observations to explore the
galaxy-wide CLFs in the starburst-ring galaxies NGC 1512 and NGC
5248. Although they do not distinguish between clusters located in the
starburst rings compared with those in the main galaxy discs, their
cluster samples are dominated by starburst ring clusters. They
separate their samples into $\le$15 Myr-old and older clusters and
show that for both galaxies the young and older extinction-corrected
$V$-band CLFs exhibit similar shapes, within the statistical
uncertainties, in essence following power-law distributions to the
lowest luminosities attainable. (A downturn of the CLFs in the
lowest-luminosity bin is apparent for NGC 5248, although the
importance of selection effects on these results is unclear since
those authors only focussed on the bright end of the CLF.)

The $V$-band CLFs of the $<$10 Myr-old cluster system in the
circumnuclear ring of the peculiar galaxy ESO 565-G11, as well as that
of the inner-ring CLF in NGC 3081 appear to be genuine power-law
functions with slopes $\alpha \simeq 2.2$ to $\alpha \simeq 2.5$ down
to the completeness limits (Buta et al. 1999, 2000a, 2004). Similarly,
Barth et al. (1995) showed that the NGC 1097 ring CLF exhibits a steep
rise down to at least the estimated observational completeness limit,
$M_V = -11.0$ mag. These slopes are reminiscent of the $\alpha = 1.8$
slope derived by de Grijs et al. (2003a) for the young, $\sim 30$
Myr-old starburst-ring clusters in NGC 3310, for $17.7 \le {\rm F606W}
\le 20.2$ mag. However, these luminosities are well above the
observational selection limit, while the NGC 3310 CLF is not shown
graphically by these authors. They do point out, however, that a
comparison of their broad-band CLF with that based on H$\alpha$ data
is inconclusive.

The star clusters in the circumnuclear rings of NGC 1326 and NGC 4314
pose interesting conundrums. Buta et al. (2000b) find that the
apparent $V$-band CLF of the NGC 1326 ring clusters, which are mostly
thought to be younger than $\sim 20$ Myr, is well approximated by a
power law down to the observations' generic completeness level of $M_V
= -8.75$ mag. However, once they apply a correction for internal
extinction, the resulting CLF exhibits a significant deviation from
the canonical power-law distribution for $M_V \ga -10.5$ mag, i.e., at
significantly brighter luminosities than the applicable completeness
limit. Unfortunately, the authors do not comment on this CLF
behaviour, which could imply that extinction might {\it mask} a CLF
turnover rather than cause it. Benedict et al. (2002) find an almost
identical result for the NGC 4314 ring clusters.

It thus appears that even in the high-density, high-shear
circumnuclear starburst rings in the local Universe the shape of the
initial CLF resembles that of a power law with the canonical slope
$\alpha = 2$. The high-density environments in these rings make them
ideal locations to harbour large numbers of YMCs, much more so than in
galaxy centres (e.g., Miocchi et al. 2006). In turn, this might make
these environments conducive to significant cluster-to-cluster
interactions. This could possibly lead to enhanced cluster disruption
(e.g., de Grijs \& Anders 2012; and references therein) or cluster
mergers (e.g., McMillan et al. 2007; Saitoh et al. 2011; Fujii et
al. 2012), depending on their relative bulk velocities. Both processes
facilitate the preferential removal of lower-mass clusters from a
cluster population, although we point out that for cluster mergers to
be a viable mechanism, the resulting mass distribution must somehow
remain scale-free. It is unclear at the present time whether this
constraint is met in practice. More importantly, however, the direct
collision rate is expected to be significantly lower than the more
distant encounter rate, even in these high-density starburst rings,
simply owing to the difference in collision cross section versus the
impact radius of more distant encounters. In fact, simulations have
shown that as soon as the relative velocities are sufficiently high
compared with the clusters' internal velocity dispersions, cluster
mergers are inefficient and destructive encounters start to dominate
(Fellhauer et al. 2009; for the basic physical principles involved,
see also Gerhard \& Fall 1983).

For example, the nuclear ring in NGC 1512 is located very close to the
galactic centre, with inner and outer edges at radii of 288 pc and 722
pc, respectively. Within this specified narrow range of radii, we
estimated an approximate cluster number density by adopting a standard
vertical scale height of $\sim$300 pc (cf. the thin disc of our Milky
Way) of 649 clusters kpc$^{-3}$, corresponding to an average
cluster-to-cluster separation of $\sim$115 pc for clusters with
luminosities above our selection limit and after correction for the
$B$-band detection limit. We adopted Galactic parameters for NGC 1512,
because (i) both galaxies are barred spirals and (ii) they have
similar dynamical masses of $\sim 5 \times 10^{11}$ M$_{\odot}$ within
roughly 60 kpc (e.g., Hartwick \& Sargent 1978; Koribalski \&
L\'opez-S\'anchez 2009). For comparison, we obtained an approximate
density of 300 massive clusters kpc$^{-3}$ in the NGC 7742 starburst
ring, or an average separation of $\sim 150$ pc (de Grijs \& Anders
2012), if we adopt Mazzucca et al.'s (2008) ring size and assume that
the ring thickness follows that of generic, late-type galactic discs,
with a ratio of scale length to scale height of order eight (e.g., de
Grijs 1998).

To place these numbers in their proper context, we also calculated the
approximate cluster density in the centre of the nearby barred spiral
galaxy Messier 83 (M83), based on relevant data fom the literature
(Chandar et al. 2010; Bastian et al. 2012), which resulted in an
estimate of 524 clusters kpc$^{-3}$. This number density is lower than
that in the NGC 1512 starburst ring but higher than that in the NGC
7742 ring. To arrive at this estimate, we assumed that all clusters in
the galaxy's central region are located in a disc-like configuration
and we used the Milky Way's thin-disc scale height, combined with the
area covered by the M83 central cluster sample. Our estimate of the
clusters' number density may well be an underestimate, because M83 is
a massive starburst galaxy which is significantly larger than NGC 1512
(Jarrett et al. 2003). Given the roughly constant scale height to
scale length ratio in spiral galaxies (e.g., de Grijs 1998; his fig. 6
for Hubble types greater than 3 or 4), it is likely that M83 will,
therefore, have a larger scale height than NGC 1512. We thus conclude
that the density of star clusters in the circumnuclear starburst ring
of NGC 1512 is much higher than that in the centres of
star-cluster-forming galaxies such as M83 (we found a similar result
for M51), although this same conclusion does not hold for all
starburst rings (e.g., NGC 7742). We recently carefully reassessed the
properties of the NGC 7742 cluster population, based on their
luminosities (an observational quantity) rather than their masses (a
derived quantity). Our preliminary analysis seems to confirm that the
galaxy's starburst-ring clusters -- and particularly the youngest
subsample -- show evidence of a turnover in the CLF well above the 90
per cent completeness limit (de Grijs \& Ma, in prep.).

\section*{Acknowledgements}

We thank Thijs Kouwenhoven and Mike Fellhauer for providing us with
helpful dynamical insights. This paper is based on observations made
with the NASA/ESA {\sl HST}, and obtained from the HLA, which is a
collaboration between the Space Telescope Science Institute
(STScI/NASA), the Space Telescope European Coordinating Facility
(ST-ECF/ESA), and the Canadian Astronomy Data Centre
(CADC/NRC/CSA). This research has also made use of NASA's Astrophysics
Data System Abstract Service. We acknowledge research support from the
National Natural Science Foundation of China (grants 11373010,
11633005 and U1631102).

\end{document}